\documentclass[sn-mathphys-num]{sn-jnl}

\usepackage[T1]{fontenc}
\usepackage[title]{appendix}%
\usepackage{textcomp}%
\usepackage{manyfoot}%
\usepackage{url}
\usepackage{booktabs}%
\usepackage{algorithmicx}%
\usepackage{algpseudocode}%
\usepackage{listings}%

\usepackage{amssymb}
\usepackage{amsmath}
\usepackage{amsthm}
\PassOptionsToPackage{hyphens}{url}
\usepackage[hyphens]{url}
\usepackage{hyperref}
\usepackage{bm}
\usepackage{enumerate}
\usepackage{fancyvrb}
\usepackage{color}
\usepackage{mathrsfs}
\usepackage{graphicx}
\usepackage{textcase}
\usepackage{comment}
\usepackage{accents}
\def\theequation{\arabic{section}.\arabic{equation}}
\usepackage[svgnames,x11names,table]{xcolor}
\usepackage{amsfonts}
\usepackage{cases}
\usepackage{braket}
\usepackage{multirow}
\usepackage{tabularx}
\usepackage[scr=boondoxo]{mathalfa}

\hypersetup{
    colorlinks=true,
    linkcolor=blue,
    urlcolor=blue,
    citecolor=blue,
    bookmarksdepth=5,
    breaklinks=true
}

\definecolor{pinegreen}{rgb}{0.0, 0.47, 0.44}

\def\nn{\nonumber}
\def\r{\rho} 
 
\def\ie{{\it i.e.}}

\makeatletter
\newsavebox\myboxA
\newsavebox\myboxB
\newlength\mylenA

\newcommand*\xoverline[2][0.75]{%
    \sbox{\myboxA}{$\m@th#2$}%
    \setbox\myboxB\null
    \ht\myboxB=\ht\myboxA%
    \dp\myboxB=\dp\myboxA%
    \wd\myboxB=#1\wd\myboxA
    \sbox\myboxB{$\m@th\overline{\copy\myboxB}$}
    \setlength\mylenA{\the\wd\myboxA}
    \addtolength\mylenA{-\the\wd\myboxB}%
    \ifdim\wd\myboxB<\wd\myboxA%
       \rlap{\hskip 0.5\mylenA\usebox\myboxB}{\usebox\myboxA}%
    \else
        \hskip -0.5\mylenA\rlap{\usebox\myboxA}{\hskip 0.5\mylenA\usebox\myboxB}%
    \fi}
\makeatother

\newcommand\smallO{
  \mathchoice
    {{\scriptstyle\mathcal{O}}}
    {{\scriptstyle\mathcal{O}}}
    {{\scriptscriptstyle\mathcal{O}}}
    {\scalebox{.7}{$\scriptscriptstyle\mathcal{O}$}}
  }

\begin{document}

\title{Covariant single-field formulation of effective cosmological bounces}


\author*[1,2]{\fnm{Marcello} \sur{Miranda}}\email{marcello.miranda@unina.it}

\affil[1]{Scuola Superiore Meridionale, Largo San Marcellino 10, 
I-80138, Napoli, Italy}

\affil[2]{Istituto Nazionale di Fisica Nucleare, Sezione di Napoli, Complesso Universitario Monte Sant'Angelo, Edificio G, Via Cinthia, I-80126, Napoli, Italy}


\abstract{
    This study explores the feasibility of an effective Friedmann equation in removing the classical Big Bang initial singularity and replacing it with a non-singular bounce occurring at a critical energy density value.
    In a spatially flat, homogeneous, and isotropic universe, the effective theory is obtained by introducing a function that is parametrically dependent on the critical energy density.
    This function measures the deviation from the benchmark theory, which is recovered as the critical energy density approaches infinity.
    Focusing on the covariant single-field formulation in viable Horndeski gravity, our analysis shows that both the effective and the benchmark theories belong to the same scalar-tensor theory, without any additional propagating degrees of freedom: the cuscuton and extended cuscuton models.}

\maketitle

\section{Introduction} \label{sec:intro}

According to the standard cosmology, our Universe originated from an initial singularity known as the Big Bang. This classical picture posits that as we approach the singularity the volume of the Universe tends to zero, while the energy density and temperature diverge, within the curvature scalars. 
However, it is widely accepted that classical singularities and divergences represent limitations in our current understanding of physics, suggesting the need for a more comprehensive theory. They may be resolved by considering the effects of quantum gravity, which becomes dominant over the other interactions in extreme scenarios.
{Another approach to resolving the initial singularity is through a classical non-singular bounce occurring at an energy density sufficiently below the Planck scale. In this case, either a stable form of stress-energy that violates the null energy condition or a modification of Einstein's General Theory of Relativity (GR) that bypasses the null convergence condition is necessary. Although this may lead to instabilities in some scenarios \cite{Rubakov:2014jja}, there are examples that have successfully avoided all known instabilities \cite{Ijjas:2016tpn, Ijjas:2016vtq, Ijjas:2018cdm,Tukhashvili:2023itb}.}

One of the most compelling frameworks suggests that quantum geometry might give rise to a short-range repulsive force in the Planck regime (otherwise negligible), thereby transforming the Big Bang singularity into a so-called quantum bounce. This is the case for theories such as Loop Quantum Cosmology (LQC), in which the bounce is described by an effective Friedmann equation~\cite{Bojowald:2001xe, Ashtekar:2011ni, Bojowald:2018sgf}. 

In this context, we are interested in studying the viability of an effective Friedmann equation, describing the dynamics of a spatially flat isotropic and homogeneous bouncing universe. The standard equation is modified by assuming a maximum energy density value, $\rho_c$, and we question whether an effective formulation can remove the classical initial singularity. Moreover, we want to investigate covariant formulations of non-singular bounces wherein the modifications are sourced by a scalar field in second-order theories of gravity (which are extensions of GR).
Therefore, the Universe is modelled by a spatially flat Friedmann--Lema\^itre--Robertson--Walker (FLRW) metric, $g_{\mu\nu}=\mbox{diag}(-1,a^2,a^2,a^2)$, where the scale factor $a$ is a function of the cosmological time $t$. Then, the \textit{overdot} indicates the time derivative. Notice that throughout the entire work, we use the reduced Planck units, $c=8\pi G=\hslash=1$.

Let us provide a brief overview of the initial singularity issue in GR.
Evaluating Einstein's field equations on an FLRW background, one obtains the (first) Friedmann, $H^2=\rho/3$, and the Raychaudhuri equation (or second Friedmann equation), ${\ddot{a}}/{a}=-\left(\rho+3p\right)/6$, with $H\equiv\dot{a}/a$ denoting the Hubble rate, and $\rho$ and $p=-2\dot{H}-3H^2$ representing the homogeneous and isotropic energy density and pressure of a perfect fluid, respectively.
The contracted Bianchi identity~\cite{Carroll:2004st} guarantees the conservation of the stress-energy tensor\footnote{It is well known that the continuity equation can be obtained by differentiating the first Friedmann equation and rewriting $\dot{H}=\ddot{a}/a-H^2$ in terms of energy density and pressure. Similarly, the second Friedmann equation can be obtained by differentiating the first Friedmann equation and using the continuity equation to rewrite $\dot{\r}$.}, which yields the continuity equation $\dot{\rho}+3H\left(\rho+p\right)=0$.
To solve the field equations, it is necessary to use an equation of state characterising the matter content. The simplest assumption (used to describe standard matter) is a linear barotropic equation of state, $p=w\rho$, where $w$ is the barotropic constant. Then, for $w\neq-1$, the continuity equation yields $\r=\r_*\left({a_*}/{a}\right)^{3(1+w)}$ where the quantities with the \textit{subscript} ‘$\ast$’ are integration constants. Substituting the above equation into the first Friedmann equation, considering the case of $w>-1$, there exists a cosmological time, $t_{BB}$, such that the scale factor vanishes and, therefore, the energy density diverges to infinity.

For instance, assuming the existence of a quantum short-range repulsive force occurring at the Planck scale could solve the initial singularity. The general conditions for the occurrence of a bounce are $H=0$ and $\ddot{a}/a>0$, taking place when $\r=\r_c$ (or, equivalently, $t=t_c$). However, the standard Friedmann equation does not satisfy these conditions. Moreover, for the bounce to occur in GR, the matter must violate the null energy condition~\cite{Capozziello:2014bqa, Ijjas:2016pad}, \ie, $\r+p>0$. Thus, one aims to modify the Friedmann equation to mimic the bounce at early cosmological time and trace back the standard matter behaviour at late cosmological time.

There are different ways and frameworks in which one can obtain modifications of the Friedmann equation leading to a bounce. 
In particular, for a flat FLRW universe, a simple effective Friedmann equation being suitable for a bouncing interpretation is achieved in LQC, $ H^2=\r\left(1-{\r}/{\r_c}\right)/3$, where the matter content satisfies the continuity equation $\dot{\rho}+3H(\r+p)=0$~\cite{Taveras:2008ke, Singh:2006im, Sami:2006wj, Bojowald:2012xy}.
The bounce takes place at $\r=\r_c$, corresponding to $H^2=0$. The quadratic $\r^2$ term is relevant only at very high energy density values, $\r\lesssim\r_c$, in the proximity of the bounce. Otherwise, for $\r\ll\r_c$, the correction is negligible and the standard Friedmann equation is recovered. The GR limit is realised for $\r_c\rightarrow\infty$, where the effective Friedmann equation is reduced to the standard one\footnote{In LQC, it is obtained that $\r_c\sim{c^5}/{G^2\hslash}$ in agreement with the quantum gravitational origin of the bounce. Then, the classical GR limit corresponds to a vanishing Planck length, {\it i.e.}, $\hslash\to0\Rightarrow\r_c\to\infty$.}. 

However, the quadratic correction is just one example of an effective Friedmann equation which can be obtained in LQC. For instance, looking for higher-order quantum corrections to the field strength~\cite{Mielczarek:2008zz}, it is possible to get a different modification with a critical density value shifted to the higher energies, while, modifying the handling of the Lorentzian term inside the Hamiltonian it is possible to obtain an asymmetric bounce scenario where the evolution of the universe is described by two different branches~\cite{Li:2018opr, Li:2018fco, Li:2019ipm}, before and after the bounce. The presence of multiple modifications to the standard Friedmann equation results from the lack of a systematic derivation from the Loop Quantum Gravity framework and quantization ambiguities~\cite{Hrycyna:2008yu, Dzierzak:2009ip, Renevey:2021tmh}.
Finally, in any LQC description, one can always obtain a modified Raychaudhuri equation by differentiating the modified Friedmann equation and using the continuity equation~\cite{Marto:2013soa, Wilson-Ewing:2015lia, Burger:2018hpz}. In the simpler case of the quadratic correction, it yields ${\ddot{a}}/{a}=-\frac{1}{6}\,(\r+3p)\left(1-2\,\frac{2\r+3p}{\r+3p}\,{\r}/{\r_c}\,\right)$.
Then, one can identify the effective energy density and the effective pressure, $\r_{\rm{(eff)}}=\r\left(1-{\r}/{\r_c}\right)$ and $p_{\rm{(eff)}}= p\left(1-2{\r}/{\r_c}\right)-{\r^2}/{\r_c}$, for which the usual continuity equation holds, $\dot{\rho}_{\rm{(eff)}}+3H(\r_{\rm{(eff)}}+p_{\rm{(eff)}})=0$.

Inspired by the depicted framework, one can make purely classical considerations to assess the feasibility of an effective Friedmann equation in removing the initial singularity through a classical non-singular bounce. 

The novel effective approach is based on the assumption that the effective theory can be written perturbatively in terms of another UV-divergent (benchmark) theory and an intrinsic energy scale, $\rho_c$. The latter represents the maximal value reachable by the energy density in the benchmark theory. The critical energy density constitutes the parameter that explicitly modifies the Friedmann equation. The effective correction must ensure the finiteness of all observables when evaluated at the critical point leading to a finite curvature value. Therefore, the benchmark theory and the effective theory only differ in the presence of a maximal energy density value.
The standard equations correspond to the limit as $\rho_c$ approaches infinity. We refer to {\em standard quantities} as parametrically independent of $\rho_c$ and are associated with the benchmark theory, denoted by the subscript ‘0’.

In Sec.~\ref{sec:scb}, we derive the necessary properties holding for a general effective modification of the Friedmann equation encoding a bouncing behaviour. The effective approach imposes some constraints on the effective model in the proximity of the bounce, implying that the standard energy density and the pressure satisfy the null energy condition.

In the context of modified theories of gravity~\cite{Nojiri:2006ri, Capozziello:2007ec, Sotiriou:2008rp, Capozziello:2011et, Clifton:2011jh, Nojiri:2017ncd, Heisenberg:2018vsk}, there are many realisations of repulsive effects generating a non-singular bounce (not necessarily associated with a quantum gravity origin) which are usually characterised by additional degrees of freedom compared to GR (see also~\cite{Brandenberger:2009yt, Brandenberger:2016vhg, Biswas:2010zk, Battefeld:2014uga, Cai:2014xxa, Cognola:2016gjy, Ijjas:2016vtq, Ijjas:2016tpn, Ijjas:2017pei, Kolevatov:2017voe, deCesare:2018cts, Mironov:2018oec, Mironov:2019qjt, Mironov:2019mye, Polarski:2021azv, Odintsov:2015zua, Nojiri:2019lqw}).
Moreover, there are different methods to reconstruct the modified theory mimicking an LQC-like bounce. Some of them are characterised by what is called the \textit{order reduction} technique~\cite{Bel:1985zz, Simon:1990ic} applied to metric theories of gravity~\cite{Sotiriou:2008ya, Terrucha:2019jpm, Barros:2019pvc, Bajardi:2020fxh, Miranda:2021oig, Ribeiro:2021gds}, and others are developed in the context of Palatini theories~\cite{Allemandi:2004yx, Olmo:2008nf, Barragan:2009sq, Delhom:2023xxp}. 
However, the novel analysis can open new prospects and criticisms of the realisation of cosmic bounces in effective theories.

In particular, we analyse the possibility of reproducing a non-singular cosmological bounce in a covariant formulation through a single classical scalar field.
The covariant formulation must encode a bouncing behaviour compatible with the effective modification to the first Friedman equation.
The scalar field is introduced into the model through modifications of the Einstein--Hilbert action associated with the k-essence model~\cite{Garriga:1999vw, Armendariz-Picon:2000ulo, Copeland:2006wr, Faraoni:2022gry} or general scalar-tensor theories~\cite{Deffayet:2010qz, Gomes:2013ema, Horndeski:1974wa, Deffayet:2011gz, Kobayashi:2011nu}. 
The modifications can be seen as effectively describing an exotic matter sector or departures from GR dominating at early cosmological time so that other matter species filling the universe can be neglected. The additional contributions to the GR action should automatically guarantee a bouncing mechanism.
This means that the effective theory $S_{(\rm eff)}$ is obtained from the benchmark one $S_{(0)}$ through a \textit{minimal modification} preserving the scalar field equation of motion, analogously to LQC equations, such that $\lim_{\r_c\to\infty}S_{\rm(eff)}=S_{(0)}$. Therefore, the benchmark theory is, in general, a smaller subclass of theories contained within the class to which the effective theory belongs, $S_{(0)}\subseteq S$. However, the request for a covariant effective bouncing behaviour imposes the benchmark theory to belong to the same scalar-tensor class of the effective theory.
For this reason, it is possible, and more convenient, to consider separately the cases of minimal and non-minimal coupled scalar fields.

It is important to stress that the results obtained from this analysis directly arise from the primary assumption of the presented approach, \textit{i.e.}, that the non-singular effective theory can be written perturbatively in terms of a benchmark theory, for which the standard equations hold, and an intrinsic energy scale, $\rho_c$. This assumption, along with the requirement of a covariant formulation for the effective theory, constrains the possible realizations of the effective cosmological bounce.

In Sec.~\ref{sec:kess}, we discuss the case of a general k-essence in describing the bounce. As a result, covariant effective bounces can be formulated only by a non-dynamical scalar field, \textit{i.e.}, with no propagating degrees of freedom, the cuscuton model~\cite{Afshordi:2006ad, Afshordi:2007yx, Afshordi:2009tt}. 
In Sec.~\ref{sec:horn}, we extend our treatment to a non-minimally coupled scalar field, showing that, also in this case, a covariant description of the effective bounce selects a non-dynamical scalar field representing a particular case of the extended cuscuton model~\cite{Bhattacharyya:2016mah, Iyonaga:2018vnu, Quintin:2019orx, Miranda:2022brj}.
We highlight some final comments in Sec.~\ref{sec:fin}.

\section{Effective cosmological bounces}\label{sec:scb}
\setcounter{equation}{0}

To introduce and study the properties of the effective cosmological bounce, we denote the quantities associated with the standard Friedmann equation by the subscript ‘0’. These quantities are parametrically independent of $\rho_c$. For example, the energy density and the pressure of the benchmark theory are $\rho_{(0)}$ and $p_{(0)}$, respectively. 

The effective Friedmann equation must approach the standard form $H^2_{(0)}=\rho_{(0)}/3$ in the benchmark limit. 
Saying that the effective UV-complete theory can be written as a perturbative expansion of another UV-divergent benchmark theory implies that the standard variables evolve according to the standard continuity equation, $\dot{\rho}_{(0)}+3H_{(0)}(\r_{(0)} + p_{(0)})=0$, as it represents the zeroth perturbative order, and the equations should indeed be satisfied order-by-order. Thus, the modified Friedmann equation we are searching for must admit the following parameterisation,
\begin{equation}\label{eq:modf}
    H^2=\frac{1}{3}\,\r_{(0)}f(x)\,,
\end{equation}
where $f$ is a generic function of the dimensionless variable $x\equiv1-\rho_{(0)}/\rho_c\in[0,1]$. Moreover, let $f$ be differentiable and positive in $(0,1)$, such that $\lim_{x\to0^+}f(x)=0^+$. Then, the benchmark limit corresponds to $x\to1$ ($\r_c\to\infty$), while the bounce is reached at the limit of $x\to0^+$ ($\r_{(0)}\to\r_c$). 

Notice that in the case of an asymmetric bounce, one would have, for instance, $f_{-}$ and $f_{+}$ representing the situation before and after the bounce, respectively. Here, we are not interested in distinguishing different branches because the following discussion must be valid for both, independently of the matching conditions.

Differentiating the left-hand side of Eq.~\eqref{eq:modf} gives
\begin{equation}\label{eq:dH2}
    \dfrac{d (H^2)}{dt} = 2 H \dot{H} = 2 H_{(0)} \sqrt{f(x)}\, \dot{H}\,,
\end{equation}
because of $H=H_{(0)}\sqrt{f(x)}\,$; while, from the right-hand side, using the standard continuity equation, one obtains
\begin{align}\label{eq:drf}
    \dfrac{1}{3}\dfrac{d\left(\rho_0 f(x)\right)}{dt} &= \dfrac{1}{3}\left[ f(x)\dot{\rho}_0- \dfrac{\rho_0}{\rho_c} f'(x)\dot{\rho}_0 \right]\nn\\[5pt]
    &=\dfrac{\dot{\rho}_0}{3}\left[ f(x)- (1-x) f'(x) \right]\nn\\[5pt]
    &= -H_0 (\rho_0+p_0)\left[ f(x)- (1-x) f'(x) \right]\,,\nn\\[3pt]
\end{align}
where the prime represents the derivative with respect to the variable $x$.
Then, solving for \(\dot{H}\), it turns out
\begin{equation}\label{eq:hdot}
    \dot{H}=-\frac{1}{2}\left(\rho_{(0)}+p_{(0)}\right) \,\frac{f(x)-(1-x) \, f'(x)}{\sqrt{f(x)}}\,,
\end{equation}
Thus, the modified Raychaudhuri equation associated with Eq.~\eqref{eq:modf} reads
\begin{align}
    \frac{\ddot{a}}{a}=\,-\frac{1}{6}&\Bigg[\,3\left(\rho_{(0)}+p_{(0)}\right) \,\frac{f(x)-(1-x) \, f'(x)}{\sqrt{f(x)}}
    -2\, \rho_{(0)} f(x)\Bigg]\,.\label{eq:raycha}
\end{align}

Analogously to LQC, it is possible to define an effective energy density and effective pressure to recast the modified Friedman in the standard form,
\begin{align}
    \r_{\rm (eff)}=&\,\rho_{(0)}f(x),\label{eq:reff}\\[7pt]
    p_{\rm{(eff)}}=&\,\left(\rho_{(0)}+p_{(0)}\right) \,\frac{f(x)-(1-x) \, f'(x)}{\sqrt{f(x)}}-\rho_{(0)} f(x),\label{eq:peff}
\end{align}
for which the continuity equation $\dot{\rho}_{\rm (eff)}+3H(\rho_{\rm (eff)}+p_{\rm (eff)})=0$ holds.
Then, the limit $x\to1$ provides the asymptotic benchmark behaviour, $\r_{\rm (eff)}\to\rho_{(0)}$ and $p_{\rm (eff)}\to p_{(0)}$.

In correspondence of $\rho_{(0)}=\rho_c$, our system must satisfy the conditions $H=0$ and $0<\ddot{a}/a<\infty$ to have a viable cosmological bounce.
Since the first condition is satisfied by definition, the only condition to impose is the second one, or equivalently, $0<\dot{H}<\infty$ at the bounce. 
In this regard, let us analyse the contribution due to the presence of \( f \) in Eq.~\eqref{eq:hdot}. Approaching the bounce,
\begin{equation}
\lim_{{x \to 0}} \frac{f(x)-(1-x)f'(x)}{\sqrt{f(x)}} = - \lim_{{x \to 0}} \frac{f'(x)}{\sqrt{f(x)}}\,.
\end{equation}
It is possible to prove that the above limit is always non-positive. Since \( f(x) \) is positive-definite on \( (0,1] \) and approaches zero as \( x \) approaches \( 0^+ \), its first derivative must be positive in the immediate proximity of zero, while it can vanish at the bounce\footnote{If \( f'(x) \) were negative in the immediate proximity of zero—meaning there is no arbitrarily small neighborhood \( (0, \epsilon] \) where \( f'(x) \) is positive—then \( f(x) \) would take on negative values, contradicting the positive-definiteness hypothesis.}, namely, \( \lim_{x \to 0^+} f'(x) \ge 0 \). Consequently, the above limit can only be zero, finite, or vanishing, corresponding to \(f'(x) \sim \smallO(\sqrt{f(x)})\), \(f'(x) \sim \sqrt{f(x)}\), and \( f'(0^+) > 0 \), respectively.

To ensure \(\dot{H}\) is finite at the bounce, we require that \( f'(x) \sim \sqrt{f(x)} \), so that 
\begin{equation}\label{eq:limit_f}
    \lim_{{x \to 0}} \frac{f'(x)}{\sqrt{f(x)}} = \ell\,,
\end{equation}
where \(\ell\) is a constant.
In Eq.~\eqref{eq:limit_f}, we are evaluating the ratio of two positive quantities in the limit as both the numerator and the denominator approach zero. This implies that the only case in which the above limit is finite corresponds to \(\ell > 0\).
Then, the defining properties of \(f\) imply that Eq.~\eqref{eq:hdot} is compatible with an effective cosmological bounce only for
\begin{equation}\label{eq:limit_l}
    f: \lim_{x \to 0^+}\frac{f'(x)}{\sqrt{f(x)}} = \ell > 0\,.
\end{equation}

Assuming \( f \) is sufficiently regular, differentiating the numerator and denominator, one obtains
\begin{align}
    \ell = &\lim_{x \to 0^+} \frac{f'(x)}{\sqrt{f(x)}} = \lim_{x \to 0^+} 2f''(x)\frac{\sqrt{f(x)}}{f'(x)} = \lim_{x \to 0^+} \frac{2}{\ell}\,f''(x)\nn\\[5pt]
    \Rightarrow\enspace &\lim_{x \to 0^+}f''(x) = \frac{\ell^2}{2} > 0\,.\label{eq:second_f}
\end{align}
Therefore, approaching the critical energy density, the conditions ensuring a viable cosmological bounce described by a general effective Friedmann equation~\eqref{eq:modf} are the following:
\begin{equation}\label{eq:asymtotics}
    \begin{cases}\\[-10pt]
        \,f(x)\sim0\,,\\[5pt]
        \,f'(x)\sim\ell\sqrt{f(x)}\,,\\[5pt]
        \,f''(x)\sim\ell^2/2\,.\\[5pt]
    \end{cases}
\end{equation}
Integrating the above asymptotic behaviour yields
\begin{equation}\label{eq:asymtotic}
    f(x)\sim\frac{\ell^2}{4} x^2\,.
\end{equation}

This analysis has an important consequence: the benchmark theory must satisfy the null energy condition at the bounce. Indeed, at the bounce, Eq.~\eqref{eq:hdot} reads
\begin{equation}
    \dot{H} = \frac{1}{2}\left(\rho_{c}+p_{c}\right) \,\ell\,,
\end{equation}
where $p_c=p_{(0)}\big|_{\substack{t=t_c}}$ is the critical value of the pressure, with $t_c$ being the time at which the bounce occurs. Knowing from the above discussion that \(\ell\) is positive, the condition \(\dot{H} > 0\) at the bounce implies that the standard matter must satisfy the null energy condition, \(\rho_{(0)} + p_{(0)} > 0\). This general result naturally comes from the request that Eq.~\eqref{eq:modf} describes a general effective cosmological bounce (admitting a perturbative approach) and holds for any benchmark theory\footnote{Notice that considering the pathological cases \( f'(0^+) > 0 \) and \(f'(x) \sim \smallO(\sqrt{f(x)})\), the sum \(\rho_{(0)} + p_{(0)}\) should properly go to zero and diverge, respectively, to guarantee the condition $\dot{H}>0$. One could include those cases by requiring \(\rho_{(0)} + p_{(0)} \ge 0\). However, they are out of our interest since it is possible to demonstrate that they do not admit a covariant single-field formulation of an effective cosmological bounce.}. It is worth stressing that, while it is well known that \(\rho_{\rm (eff)}\) and \(p_{\rm (eff)}\) must violate the null energy condition to have a bounce, here, the standard variables are the ones that must satisfy it (at the bounce).

From Eq.~\eqref{eq:asymtotic}, a class of regular solutions is characterized by $f(x)=h(x)\,x^2$ where $h(x)$ is a regular positive-definite function\footnote{If $f(x)$ is analytic, it is always possible to write $f(x)=h(x)x^2$ where $f(x)=\sum_{n=0}^{\infty}{f^{(n+2)}(0)}x^n/{(n+2)!}$.} in $x\in [0,1]$ such that $\ell= 2 \sqrt{h(0)}$.
The simplest case is for $h$ to be constant. In particular, let us consider the case $h(x)=1$ corresponding to $f(x)=x^2$ and $\ell=2$. Then, effective energy density and pressure turn out to be
\begin{align}
    \r_{\rm (eff)}=&\,\rho_{(0)}\left(1-\frac{\rho_{(0)}}{\rho_c}\right)^2\,,\\[7pt]
    p_{\rm (eff)}=&\,(\r_{(0)}+p_{(0)})\left(1-3\frac{\rho_{(0)}}{\r_c}\right)-\,\rho_{(0)}\left(1-\frac{\rho_{(0)}}{\rho_c}\right)^2\,,\nn\\[4pt]
\end{align}
and the modified Friedmann equations read
\begin{align}
    H^2=&\,\frac{1}{3}\,\rho_{(0)}\left(1-\frac{\rho_{(0)}}{\rho_c}\right)^2,\label{eq:scb1}\\[7pt]
    \frac{\ddot{a}}{a}=&-\frac{1}{6}\,\Bigg[\,3\left(\rho_{(0)}+p_{(0)}\right)\left(1-3\frac{\rho_{(0)}}{\r_c}\right)
    -2\,\r_{(0)}\left(1-\frac{\rho_{(0)}}{\rho_c}\right)^2\Bigg]\,.\label{eq:scb2}
\end{align}
The above one is the simplest effective cosmological bounce realisation. In the limit for $\r_c\to\infty$ it reproduces the standard Friedmann equations, and for $\r_{(0)}\to\r_c$ it turns out $\r_{\rm (eff)}=0$ and $p_{\rm (eff)}=-2\left(\r_{c}+p_{c}\right)$. Then, the condition $\left(\r_{c}+p_{c}\right)>0$ guarantees the bounce, associated with a negative effective pressure. 

From Eq.~\eqref{eq:scb1}, the effective Hubble rate can be written as follows,
\begin{align}
    H=&\,H_{(0)}\left(1-\frac{3H_{(0)}^2}{\r_c}\right)\,.
\end{align}
Then, assuming a linear barotropic equation of state for the standard matter, $p_{(0)}=w\r_{(0)}$, one takes back the usual GR results: {\it i.e.}, for $w\neq-1$, according to the continuity equation, it is found that
\begin{align}
   &\,\dot{\r}_{(0)}+3\frac{\dot{a}_{(0)}}{a_{(0)}}(1+w)\r_{(0)}=0
   \quad
   \Rightarrow\quad\r_{(0)}=\r_{*}\left(\frac{a_{*}}{a_{(0)}}\right)^{3(1+w)}\,,
\end{align}
and, from the standard Friedmann equation, the scale factor reads
\begin{equation}
    a_{(0)}=a_{*}\left[\,1+\frac{3}{2}(w+1)H_{*}\,(t-t_{*})\,\right]^{\tfrac{2}{3(w+1)}}\,.
\end{equation}
Finally, in terms of the shifted time $\tau=t-t_{*}+\tfrac{2}{3H_{*} (w+1)}$, the benchmark energy density and the Hubble rate turn out to be
\begin{align}
    \rho_{(0)}=&\,\frac{4}{3  \tau^2 (w+1)^2}\,,\label{eq:rho0}\\[5pt]
    H=&\,\frac{2}{3}\frac{1}{ \tau (w+1)}\left(1-\frac{4}{3  \tau^2 (w+1)^2\r_c}\right)\,,
\end{align}
showing that the bounce takes place at $\tau=\tau_c$, where
\begin{equation}
    \tau_c =\frac{2}{\sqrt{3\r_c} }\frac{1}{(w+1)}\,,
\end{equation}
which approaches zero, the classical initial singularity, in the limit $\r_c\to\infty$.

It is important to stress that the effective model~\eqref{eq:scb1} is used just as an example, being one of the simplest analytical solutions of an effective viable bounce within the framework under consideration. However, even if this particular choice ($h=1\Rightarrow \ell=2$) lacks a compelling physical motivation, it encodes the asymptotic behaviour~\eqref{eq:asymtotic}. Looking for $f$ being a third-order polynomial, the simplest realisation generalising the previous example is $h(x)= x+(1-x) \ell ^2/4$.

{The effective bounce model inherits stability features from the asymptotic benchmark theory. The perturbation of the effective part decays as the universe moves away from the bounce, indicating that the bouncing behaviour is stable against small perturbations. The easiest way to check the stability property is by perturbing the system $\rho_{(0)}\to \rho_{(0)}+\delta\rho_{(0)}$ close to the solution~\eqref{eq:rho0} and verifying that the perturbed equations describe a non-singular bounce in correspondence with $\tau=\tau_c+\delta\tau$.}

The presented framework is also compatible with LQC formulations. One needs to remember the only assumption on which the entire discussion is based: the effective bouncing theory can be written in terms of a UV-diverging benchmark theory. Therefore, for instance, the effective Friedmann equation $H^{2}=\rho\,(1-\rho/\rho_c)/3$ must be rewritten to make explicit $\rho_{(0)}$ and $\rho_c$ dependences. Indeed, the energy density $\rho$ does parametrically depend on the critical energy because it is associated with the continuity equation $\dot{\rho}+3H(\rho+p)=0$ where $H=H_{(0)}\sqrt{f(x)}$, in our framework. This implies that LQC energy density must admit a parameterisation of the following form,
\begin{equation}\label{eq:rho}
    \rho=\rho_{(0)}\,\varepsilon(x)\,,
\end{equation}
where $\varepsilon(x)$ is a positive-definite continuous function in $[0,1]$, with $\lim_{x\to1^-}\varepsilon(x)=1$, to obtain the right standard limit, and $\lim_{x\to0^+}\varepsilon(x)=1$, such that $\rho_c$ is the maximal value of energy density\footnote{This condition can be relaxed by assuming that $\lim_{x\to x_{c}^{+}}\varepsilon(x)>0$ is finite, so that $\rho=\rho_{(0)}\,\varepsilon\to\rho_c$ corresponds to the bounce. Therefore, as $\rho$ approaches $\rho_c$, the variable $x=1-\rho_{(0)}/\rho_c$ approaches $x_c=1-\rho_{(0)c}/\rho_c$, where $\rho_{(0)c}$ and $x_c$ represent the maximum values of $\rho_{(0)}$ and $x\in[0,x_c]$, respectively, and $\varepsilon(x)$ tends to $\rho_c/\rho_{(0)c}$. The final result does not change with respect to our purpose.}.
Then, in the case of the quadratic correction, $f(x)=\varepsilon(x)\left[\,1-(1-x)\,\varepsilon(x)\,\right]$, imposing the effective cosmological bounce conditions~\eqref{eq:asymtotics} and~\eqref{eq:asymtotic}, it turns out that $\lim_{x\to0^+}\varepsilon'(x)=1$ and $\lim_{x\to0^+}\varepsilon''(x)=2-{\ell ^2}/{2}$. 
To give an example, assuming $\ell=2$, a simple analytical realisation of the quadratic LQC correction is $\varepsilon(x)=1+x-x^3$ which corresponds to $f(x)=x^7-x^6-2 x^5+2 x^3+x^2$. Keeping general the value of $\ell$, the simplest case is $\varepsilon(x)=1- (1-x) x \left[x \left(\ell ^2-8\right)-4\right]/4$.

Finally, using the continuity equation $\dot{\rho}+3H(\rho+p)=0$, it is possible to obtain the expression of the pressure $p$ associated with $\rho$, which, at the bounce, reads as $p\big|_{\substack{t=t_c}}=\ell\,p_c+\left(\ell-1\right) \rho_c$. Consequently, at the bounce, $\rho+p=\ell\left(\rho_c+p_c\right)>0$, meaning that LQC variables also satisfy the null energy condition in this framework.

\section{Minimally coupled scalar field}\label{sec:kess}
\setcounter{equation}{0}

In this section, we study the possibility of implementing a general effective cosmological bounce in a covariant way, by taking into account a general k-essence model~\cite{Copeland:2006wr},
\begin{equation}\label{eq:kess}
    S=\int{d^4x\sqrt{-g}\left[\frac{1}{2}R+{L}(\phi, X)\right]}\,,
\end{equation}
with $X=-\frac{1}{2}\nabla^{\mu}\phi\nabla_{\mu}\phi$ and $\sqrt{-g}$ being the square root of the metric determinant.

The field equations of the above theory are given by
\begin{equation}
G_{\mu\nu}=L_{,X}\nabla_{\mu}\phi\nabla_{\nu}\phi+L\, g_{\mu\nu}\,,\label{eq:kfeq}
\end{equation}
coming from the variation of~\eqref{eq:kess} with respect to the metric tensor, while the variation of the action relative to the scalar field gives
\begin{align}
\left( {L}_{,X} g^{\alpha\beta} - 
{L}_{,XX} \nabla^\alpha \phi \nabla^\beta \phi 
\right)\nabla_\alpha\nabla_\beta \phi- 2 X {L}_{,X \phi} + {L}_{,\phi}&= 0 \,,\label{eq:keom}
\end{align}
where $L_{, \phi}\equiv\partial L/\partial \phi$ and $L_{, X}\equiv\partial L/\partial X$.

For a spatially flat FLRW spacetime with a homogeneous scalar field, $\phi=\phi(t)$ and $X=\frac{1}{2}\dot{\phi}^2$, the scalar field is described by a perfect fluid stress-energy tensor~\cite{Faraoni:2022gry}. It is possible to interpret the right-hand side of Eq.~\eqref{eq:kfeq} as the stress-energy tensor, $T_{\mu\nu}^{(\phi)}$, associated with the scalar field,
\begin{equation}
    T_{\mu\nu}^{(\phi)}= \r^{(\phi)} \,u_{\mu}u_{\nu}+p^{(\phi)}\,h_{\mu\nu}\,,
\end{equation}
where $h_{\mu\nu}=g_{\mu\nu}+u_{\mu}u_{\nu}$, and
\begin{equation}\label{eq:kess_eff}
    \rho^{(\phi)}=2X\,L_{,X} -L \,,\qquad p^{(\phi)}= L\,.
\end{equation}
with $u^{\mu}=\nabla^{\mu}\phi/\sqrt{2X}=-\,\mbox{Sign}(\dot{\phi})\,\delta^{\mu}_{t}$.
 Hence, the partial differential equation~(\ref{eq:keom}) is hyperbolic and $\phi$ describes a propagating degree of freedom if the inequality ${L}_{,X} + 2 X {L}_{,XX} > 0$ is satisfied. The equation of motion for the scalar field is linked to the conservation of the stress-energy tensor; indeed, the covariant derivative of the scalar field stress-energy tensor is proportional to the gradient of the scalar field through the left-hand side of Eq.~\eqref{eq:keom}, {\it i.e.}, $\nabla^{\mu}T_{\mu\nu}^{(\phi)}=-\tfrac{1}{2}(\delta L/\delta\phi)\nabla_{\nu}\phi$ which is vanishing on-shell.

In order to model the effective bounce through $L(\phi, X)$ representing the effective theory, let $\phi$ be a scalar field independent of $\r_c$ and associated with $L_{(0)}(\phi, X)$, the asymptotic benchmark theory\footnote{We can always write the effective theory in terms of the standard variables.}.
The quantities $\rho^{(\phi)}$ and $p^{(\phi)}$ correspond to the effective energy density and pressure in Eqs.~\eqref{eq:reff} and~\eqref{eq:peff}, respectively. The standard quantities are associated with the Lagrangian $L_{(0)}(\phi, X)$ and read as follows,
\begin{equation}\label{eq:kess_variables}
    \rho_{(0)}=2X\,L_{(0),X} -L_{(0)}\,,\qquad p_{(0)}= L_{(0)}\,.
\end{equation}
Therefore, we investigate the functional form of $L(\phi, X)$ such that $\lim_{\r_c\to\infty}L(\phi,X)=L_{(0)}(\phi,X)$ and leading to an effective cosmological bounce. Thus, it is necessary to impose that the energy density and pressure~\eqref{eq:kess_eff} behave as the effective energy density and pressure, while~\eqref{eq:kess_variables} are the standard one.

Using the above standard quantities, from Eqs.~\eqref{eq:reff}, \eqref{eq:peff} and~\eqref{eq:kess_eff}, one can derive the following equations,
\begin{gather}
    2X\,L_{,X} -L=\,\rho_{(0)}f(x)\,,\label{eq:rhof}\\
    \nn\\
    L=2X\,L_{(0),X}\,\frac{f(x)-(1-x) \, f'(x)}{\sqrt{f(x)}}-\rho_{(0)} f(x),\label{eq:pf}
\end{gather}
where $x=1-\rho_{(0)}/\r_c$ and $2X\,L_{(0),X}=\rho_{(0)}+p_{(0)}$\,. Now, substituting Eq.~\eqref{eq:pf} into Eq.~\eqref{eq:rhof} yields
\begin{equation}
    L_{,X}=L_{(0),X}\,\frac{f(x)-(1-x) \, f'(x)}{\sqrt{f(x)}}\,.
\end{equation}
The above equation must be equal to the derivative with respect to $X$ of Eq.~\eqref{eq:pf}, giving a consistency equation which reads as follows
\begin{gather}
    \left\{\frac{X\,L_{(0),X}}{\r_c}\,\frac{ \left( x-1 \right) f'(x)^2-f(x) \big[\,2 \left( x-1 \right) f''(x)+3 f'(x)\,\big]}{f(x)^{3/2}}\right.\nn\\[3pt]
    \left.-\frac{\left(x-1\right) \,f'(x)+f(x) }{\sqrt{f(x)}/(\sqrt{f(x)}-1)}\right\}\Big(L_{(0),X}+2X\,L_{(0),XX}\Big)=0.\label{eq:consisency}
\end{gather}
Apparently, the consistency equation seems to constrain both $L_{(0)}$ and $f$. It is satisfied if and only if one of the two brackets is vanishing.

In particular, the case of
\begin{equation}\label{eq:cuscutonpde}
    L_{(0),X}+2X\,L_{(0),XX}=0\,,
\end{equation}
corresponds to
\begin{equation}\label{eq:cuscuton}
    L_{(0)}= \mu(\phi)\,\sqrt{2X}-V(\phi)\,,
\end{equation}
where $\mu$ and $V$ are two arbitrary functions of the scalar field, representing a non-dynamical scalar field usually called \textit{cuscuton}\footnote{It is worth highlighting that the cuscuton model is not dynamical only on a cosmological background where the scalar field is assumed to have a time-like gradient (see also~\cite{Miranda:2022wkz}).}~\cite{Afshordi:2006ad, Afshordi:2007yx, Afshordi:2009tt, Bhattacharyya:2016mah, Iyonaga:2018vnu, Quintin:2019orx, Miranda:2022brj}.

The first curly bracket in Eq.~\eqref{eq:consisency} can be studied only if {the first term, ${X\,L_{(0),X}}/{\rho_c}$, can be expressed in terms of $x$. Then, imposing that the quantity inside the curl brackets is vanishing yields a differential equation for $f(x)$.  So, noticing that $L_{(0)}$ is associated with the benchmark theory and does not depend on $\rho_c$, and that $x\equiv1-\rho_{(0)}/\rho_c$, it must hold ${X\,L_{(0),X}}/{\rho_c}\propto{\rho_{(0)}}/{\rho_c}= 1-x$, due to the presence of $\rho_c$ in the denominator. Consequently, since ${2X\,L_{(0),X}}=\rho_{(0)}+p_{(0)}$, one is requiring $\rho_{(0)}+p_{(0)}\propto\rho_{(0)}$ which is equivalent to $p_{(0)}=w\rho_{(0)}$, with $w$ being a constant, corresponding to $L_{(0)}=w(2X L_{(0),X}-L_{(0)})$}. Therefore, the benchmark theory must satisfy the following differential equation:
\begin{equation}\label{eq:w}
    \left(1+w\right) L_{(0)} - 2 w X L_{(0),X}=0\,,
\end{equation}
where $w\neq-1$ is constant. In this way, Eq.~\eqref{eq:consisency} is a well-defined differential equation for $f(x)$.  

Solving Eq.~\eqref{eq:w}, it turns out
\begin{equation}\label{eq:w_model}
    L_{(0)}=\mu(\phi)\, X^{\,\tfrac{w+1 }{2 w}}\,.
\end{equation}
In this case, the scalar field is associated with the barotropic equation of state $p_{(0)}=w\rho_{(0)}$, where $w$ must be greater than $-1$ to guarantee $\r_{(0)}+p_{(0)}>0$ (a necessary condition for the bounce occurrence as shown in the previous section)\footnote{For $w=1$ and redefining $\phi$, Eq.~\eqref{eq:w_model} can be reduced to the standard massless scalar field.}.
Then the differential equation for $f$ reads as follows,
\begin{gather}
    (w+1)(1-x)^2\frac{  f'(x)^2-f(x) \big[2  f''(x)+3 f'(x)/(1-x)\big]}{f(x)^{3/2}}\nn\\[5pt]
    +2\left(\sqrt{f(x)}-1\right)\frac{(1-x) \,f'(x)+f(x) }{\sqrt{f(x)}}=0\,.\label{eq:ode_f}
\end{gather}    
The above differential equation must be defined in the entire interval $x\in[0,1]$, and, in particular, in correspondence with the bounce. Evaluating the above equation for $x$ approaching $0^+$, according to Eqs.~\eqref{eq:asymtotics} and~\eqref{eq:asymtotic}, one obtains
\begin{equation}
    \left({3w +5}\right) \,\ell =0 \quad\Rightarrow\quad w=-\frac{5}{3}\quad\Rightarrow\quad \rho_{(0)}+p_{(0)}<0\,. 
\end{equation}
Thus, the Lagrangian~\eqref{eq:w_model} cannot describe any effective bounce because it violates the null energy condition necessary for removing the singularity in Eq.~\eqref{eq:hdot}. Then, only the cuscuton field can provide a covariant effective formulation. Then, the limiting curvature mechanism is realized by the Lagrangian~\eqref{eq:pf} where $\rho_{(0)}$ and $L_{(0)}$ are given respectively by Eqs.~\eqref{eq:kess_variables} and~\eqref{eq:cuscuton}, respectively.

\section{Non-minimal coupling case}\label{sec:horn}
\setcounter{equation}{0}

We can repeat a similar analysis to the one in the previous section, taking into account the viable (or reduced) Horndeski gravity~\cite{Horndeski:1974wa, Deffayet:2011gz, Kobayashi:2011nu}, where the adjective “viable” is used to indicate Horndeski subclass in which tensor perturbations propagate at the speed of light~\cite{Creminelli:2017sry, Baker:2017hug, Bettoni:2016mij, Andreou:2019ikc}. 

Using the modern notation of the Horndeski functions, the action reads as follows
\begin{equation}
    S=\frac{1}{2}\int{d^4x\sqrt{-g}\left[G_{4}(\phi)R+G_{2}(\phi,X)-G_{3}(\phi,X)\Box \phi\right]}\,,\label{eq:horn}
\end{equation}
where $\Box\phi=\nabla_{\alpha}\nabla^{\alpha}\phi$, the function $G_2$ represents the k-essence contribution, $G_3$ is the kinetical braiding term, and $G_4$ is the non-minimal coupling function. The field equations associated with the action~\eqref{eq:horn} can be written in terms of effective Einstein equations, $ G_{\alpha\beta}=T^{(\phi)}_{\alpha\beta}$, where the scalar field stress-energy tensor is given by the sum of the individual contributions associated with $G_{i}$ functions,
\begin{align}
      T^{(2)}_{\alpha\beta}=\,&\frac{1}{2G_4} \left( G_{2,X}\nabla_{\alpha}\phi\nabla_{\beta}\phi+G_{2}\,g_{\alpha \beta} \right)\,,\\
    &\nn\\
       T^{(3)}_{\alpha \beta}=&\frac{\left( G_{3,X} \nabla_{\gamma} X  \nabla^{\gamma} \phi  - 2 X G_{3,\phi} \right)}{2G_4}  g_{\alpha \beta} - \frac{G_{3,X}}{G_4} \nabla_{(\alpha} X \nabla_{\beta)} \phi\nn\\
    \,& - \frac{\left( 2 G_{3,\phi} + G_{3,X} \Box \phi \right)}{2G_4} \nabla _\alpha  \phi \nabla_\beta \phi\\
    &\nn\\
       T^{(4)}_{\alpha \beta}=\,&\frac{G_{4,\phi}}{G_4}(\nabla_{\alpha}\nabla_{\beta}\phi-g_{\alpha \beta}\Box\phi)
       +\frac{G_{4,\phi\phi}}{G_4}(\nabla_{\alpha}\phi\nabla_{\beta}\phi+2X\,g_{\alpha \beta})\,.
\end{align}
where, parentheses encompassing indices indicate the symmetrization, \ie, $V_{(\alpha\beta)}=(V_{\alpha\beta}+V_{\beta\alpha})/2$.
In addition, the equation of motion of the scalar field guarantees the conservation of the effective stress-energy tensor, $\nabla^{\nu}T^{( \phi)}_{\mu\nu}=-\tfrac{1}{2}(\delta L/\delta\phi)\nabla_{\mu}\phi \big|_{\rm on-shell} = 0$.

Because of the presence of $G_4$ and $G_3$, the effective tensor $T^{(\phi)}_{\mu\nu}$ behaves as an imperfect fluid~\cite{Pimentel:1989bm, Deffayet:2010qz, Pujolas:2011he, Faraoni:2018qdr} characterised by dissipative (viscous) contributions inside the effective energy density and the effective pressure~\cite{Giusti:2021sku, Miranda:2022wkz, Miranda:2024dhw}. Specialising the discussion to the spatially flat FLRW geometry, the energy density and the pressure associated with the scalar field read as follows,
\begin{equation}\label{eq:horn_eff}
\begin{aligned}
    &\rho^{(\phi)}=\bar{\rho}+3H\xi\,, \\[7pt]
    &p^{(\phi)}=\bar{p}-3H\zeta+\xi\,\ddot{\phi}/\sqrt{2X}\,
\end{aligned}
\end{equation}
where, 
\begin{align}
    \bar{\rho}=&\,\frac{\left(2XG_{2,X} -G_2-2XG_{3,\phi}\right)}{2G_{4}}\,,\\[3pt]
    \bar{p}=&\,\frac{\left({G_2}-2 X {G_{3,\phi}}+4 X {G_{4,\phi\phi}}\right)}{2 {G_4}}\,,\\[3pt]
    \xi=\,&\frac{\sqrt{2X}\left(G_{4,\phi}-X  G_{3,X} \right)}{G_{4}}\,,\\[3pt]
    \zeta=&\,\frac{2}{3}\frac{\sqrt{2X}\,G_{4,\phi}}{G_{4}}\,.\zeta
\end{align}

Analogously to the previous section, we want to use the functions $G_{i}$ to model a general effective cosmological bounce. Therefore, Eq.~\eqref{eq:horn_eff} corresponds to the effective quantities in Eqs.~\eqref{eq:reff} and~\eqref{eq:peff}. Then, the standard energy density read
\begin{equation}\label{eq:horn_variables}
\begin{aligned}
    &\rho_{(0)}=\,\bar{\rho}_{(0)}+3H_{(0)}\xi_{(0)}\,,\\[7pt]
    &p_{(0)}=\,\bar{p}_{(0)}-3H_{(0)}\zeta_{(0)}+\xi_{(0)}\,\ddot{\phi}/\sqrt{2X}\,
\end{aligned}
\end{equation}
but with the subscript ‘0’ meaning that they are associated with the benchmark functions $G_{(0)i}=\lim_{\rho_c\to\infty}G_{i}$. Using the above quantities, and imposing Eqs.~\eqref{eq:horn_eff} and~\eqref{eq:horn_variables} satisfy Eqs.~\eqref{eq:reff} and~\eqref{eq:peff}, it is possible to obtain the functional form of the effective theory.

At this point, the analysis can be performed through two alternative ways, bringing us to the same conclusion: requiring that the different contributions to effective and standard quantities are connected according to Eqs.~\eqref{eq:reff} and~\eqref{eq:peff}; rewriting the $\ddot{\phi}$ contribution in terms of the Hubble rate and Horndeski functions by using the field equation of the scalar field~\cite{Miranda:2022wkz}. In both cases, it turns out that
\begin{align}
    \xi=0&\quad\Rightarrow\quad G_{3}=G_{4,\phi}\,\ln{\xoverline{X}}\,,\label{eq:G3}\\[7pt]
    \xi_{(0)}=0&\quad\Rightarrow\quad G_{(0)3}=G_{(0)4,\phi}\,\ln{\xoverline{X}}\,,\label{eq:G03}
\end{align}
where $\xoverline{X}=X/X_{*}$ is used to make the logarithm argument dimensionless\footnote{We are not considering pure scalar field contributions in $G_{3}$ since they can be eliminated by integrating by parts, namely $-F(\phi) \, \Box\phi=2 X\,F_{,\phi}(\phi) $ up to a total derivative. Therefore, $G_{3}=F(\phi)$ is equivalent to considering $\widetilde{G}_3=0$ and $\widetilde{G}_2=G_{2}+2X\,F_{\phi}$.}, and,
\begin{equation}\label{eq:G4}
    \frac{G_{4,\phi}}{G_{4}}=\frac{G_{(0)4,\phi}}{G_{(0)4}}\left(\frac{f(x)-(1-x) \, f'(x)}{f(x)}\right)\,.
\end{equation}

{To clarify this point, it is essential to keep in mind that we are seeking a covariant formulation of an effective cosmological bounce. Consequently, we cannot impose any condition in addition to Eqs.\eqref{eq:reff} and~\eqref{eq:peff} that does not arise directly from the field equations. This requires us to eliminate the explicit presence of the second time derivative of the scalar field, yielding $\xi=0$ and $\xi_{(0)}=0$, from which Eqs.~\eqref{eq:G3} and~\eqref{eq:G03} hold. For the same reason, $\zeta_{(0)}$ and $\zeta$ must be related to each other analogously to the relation between $p_{(0)}$ and $p$, i.e.,
\begin{equation}
    \zeta=\zeta_{(0)}\left(\frac{f(x)-(1-x) \, f'(x)}{f(x)}\right)\,,
\end{equation}
which, substituting the expressions of $\zeta$ and $\zeta_{(0)}$, corresponds to Eq.~\eqref{eq:G4}.}

{Since $G_{4}$ and $G_{(0)4}$ are only functions of the scalar field, Eq.~\eqref{eq:G4} implies that $x$ must be independent of the kinetic term $X$, namely, $\rho_{(0)}$ depends only on the scalar field. Noticing that $\rho_{(0)}$ turns out to be
\begin{equation}
    \rho_{(0)}=\frac{\left(2XG_{(0)2,X} -G_{(0)2}-2X G_{(0)4,\phi\phi}\ln{\bar{X}}\right)}{2G_{(0)4}}\,,
\end{equation}
imposing the condition $\rho_{(0),X}=0$ yields the following differential equation,}
\begin{equation}\label{eq:condition_G2}
    G_{(0)2,X}+2X G_{(0)2,XX}=2\left(1+\ln{\bar{X}}\right)G_{(0)4,\phi\phi}\,,
\end{equation}
which admits the following general solution
\begin{align}\label{eq:G02}
    G_{(0)2}=\mu_{(0)}(\phi)\sqrt{2X}-V_{(0)}(\phi)+2 X  \left(\ln{\bar{X}}-2\right)G_{(0)4,\phi\phi}\,.
\end{align}
{Then, imposing that
\begin{equation}
    \bar{p}=\bar{p}_{(0)}\left(\frac{f(x)-(1-x) \, f'(x)}{f(x)}\right)\,,
\end{equation}
substituting Eq.~\eqref{eq:G02} implies the UV-complete theory is described by the following functional form,}
\begin{equation}
    G_{2}=\mu_{}(\phi)\sqrt{2X}-V_{}(\phi)+2 X  \left(\ln{\bar{X}}-2\right)G_{4,\phi\phi}\,,
\end{equation}
where,
\begin{align}
    \mu(\phi)=&\,\mu_{(0)}(\phi)\frac{G_{4}}{G_{(0)4}}\frac{f(x)-(1-x) \, f'(x)}{\sqrt{f(x)}}\,,\\[7pt]
    V(\phi)=&\,V_{(0)}(\phi)\frac{G_{4}}{G_{(0)4}}f(x)\,,
\end{align}
while $G_{4}$ is written in terms of the benchmark theory according to Eq.~\eqref{eq:G4}.

Eqs.~\eqref{eq:G02} and~\eqref{eq:G03} correspond to a subclass\footnote{Also assuming from the beginning an extended cuscuton model, it turns out that $\rho_{(0)}$ must not depend on the kinetic term. This condition corresponds to Eqs.~\eqref{eq:G03} and~\eqref{eq:G02}.} of the extended cuscuton formulation~\cite{Bhattacharyya:2016mah, Iyonaga:2018vnu, Quintin:2019orx, Miranda:2022brj}. Therefore, also in the case of viable Horndeski, the effective cosmological bounce selects a non-dynamical scalar field. It is straightforward to verify that for the coupling function approaching a constant value, one traces back the result of the previous section\footnote{It is worth noting that, also in the case of the cuscuton~\eqref{eq:cuscuton} the energy density does not depend on the kinetic term. In fact, $\rho_{(0),X}=0$ corresponds to Eq.~\eqref{eq:cuscutonpde}.}.

\section{Final discussion and conclusions}\label{sec:fin}

Considering a spatially flat, homogeneous, and isotropic universe, and assuming the existence of a critical energy density $\r_c$, we analyse the conditions characterising a general effective cosmological bounce. 

In the used approach, the effective Friedmann equation takes the form of Eq.~\eqref{eq:modf}, where the standard energy density $\rho_{(0)}$ is parametrically independent of $\r_c$, while the effective modification $f$ is a function of $x=1-\rho_{(0)}/\rho_c$. From Eq.~\eqref{eq:modf}, using the standard continuity equation of the benchmark theory allows us to obtain a modified Raychaudhuri equation~\eqref{eq:raycha}. Although the modified Friedmann equation does not diverge at $\rho_{(0)}=\rho_c$ by definition, this is not guaranteed for the modified Raychaudhuri equation. Therefore, it is necessary to impose the bounce condition $\ddot{a}/a>0$ (equivalent to $\dot{H}>0$) in correspondence with the critical energy density, as shown in Eq.~\eqref{eq:limit_l}.
Consequently, the condition for a viable effective cosmological bounce takes the form of Eq.~\eqref{eq:asymtotic}. 
Moreover, since the limit~\eqref{eq:limit_l} is positive, this framework imposes the benchmark theory to satisfy the null energy condition, $\rho_{(0)}+p_{(0)}>0$, at the bounce (while the effective quantities violate it).
Eq.~\eqref{eq:scb1} is the simplest analytical example of an effective cosmological bounce reproducing the required behaviour~\eqref{eq:asymtotic}. {Then, the bounce model is such that the effective theory inherits stability features from the asymptotic benchmark theory. However, at this level, it is not possible to make any additional claims about $f(x)$ beyond the constraints on the limit values of its derivatives approaching the bounce.}
The global modification to the Friedmann equation must adhere to some well-motivated non-singular bounce or a quantum theory, as in the case of LQC. As shown in Eq.\eqref{eq:rho}, LQC quantities can be contextualised in this framework by explicitly characterising the roles of standard variables and effective corrections.

We moved then to analyse covariant realisations of a general effective bounce in second-order theories characterised by an additional classical scalar field. This is done by searching for two actions describing the effective theory $S_{(\rm eff)}$ (depending on $\rho_c$) and the asymptotic benchmark $S_{(0)}$. The effective theory must automatically guarantee an effective bouncing behaviour and must approach the asymptotic one in the limit of $\r_c\to\infty$. Distinguishing between the effective and the benchmark theory, it is possible to impose the conditions~\eqref{eq:reff},~\eqref{eq:peff}, and~\eqref{eq:limit_f}, holding for a viable bounce.

First, we focused on the compatibility of an effective bounce with a general k-essence model. We demonstrated that the only possibility for the scalar field is the cuscuton. Thus, a dynamical k-essence model is incompatible with any bouncing model. This is because a dynamical model would have a benchmark limit violating the null energy condition, required for the bounce.
In fact, only the theory~\eqref{eq:w_model} yields a well-posed consistency equation~\eqref{eq:consisency}. It is characterised by the barotropic equation of state $p_{(0)}=w\rho_{(0)}$, where $w$ must be greater than $-1$ to ensure $\rho_{(0)}+p_{(0)}>0$. However, the consistency equation provides $w=-5/3$ which implies the standard matter violates the null energy condition at the bounce. 
Next, in the last section, we reiterated a similar discussion for a general viable Horndeski theory. Even in this case, the effective bouncing behaviour constrains the theory in a non-dynamical scalar field described by Eqs.\eqref{eq:G03} and~\eqref{eq:G02} representing a particular case of
the extended cuscuton model. 

The novel effective approach offers an innovative perspective to investigate bounce corrections. It opens up possibilities for studying theories that can reproduce or mimic effective cosmological bounces and exploring the limits and physical conditions required for their realisation. At this point, one could extend the discussion to more general scalar-tensor theories, but the resulting theory would always be associated with a non-dynamical scalar field. The reason is that the effective theory and the benchmark limit must reproduce the same scalar field equation of motion (associated with the continuity equation of the standard variables). This strongly constrains $S_{(\rm eff)}$ and $S_{(0)}$. 
Nevertheless, it is interesting to notice that not all the theories with no extra dynamical degrees of freedom (in addition to the two tensor modes) are compatible with an effective bounce. Moreover, adding an explicit standard matter contribution to the current framework, the conclusion would likely be the same because the standard continuity equation implies the conservation of the effective quantities, and \textit{vice versa}. In this sense, the adopted approach preserves the number of total degrees of freedom, and, in the specific case of an additional single scalar field, it selects among theories having only two tensor modes~\cite{Gao:2019twq}. 
Remarkably, these findings align with prior research. In Ref.~\cite{Boruah:2018pvq}, it has been shown that the cuscuton model allows for an effective violation of the null energy condition in FLRW backgrounds while the standard matter sources satisfy it. Studies on the cuscuton power spectrum and the absence of instabilities can be found in Refs.~\cite{Ito:2019fie, Kim:2020iwq, Bartolo:2021wpt} (see also Refs.~\cite{Papallo:2017qvl, Kovacs:2020ywu}).

In this work, we focused only on scalar-tensor theories. Other modified theories of gravity would need a different \textit{ad hoc} analysis. It would be interesting to specialise this approach to alternative theories of gravity. Moreover, one can generalise the effective bounce to different cosmologies than FLRW. We leave this discussion to the near future, addressing this issue in forthcoming works. 

\backmatter

\bmhead{Acknowledgements}

The author thanks Marco Pozzetta for helping to see things from a different perspective, Marco De Cesare for his insightful feedback and valuable suggestions, Salvatore Capozziello, Nicola Menadeo, and Daniele Vernieri for their engaging discussions, and acknowledges the support of the Istituto Nazionale di Fisica Nucleare (INFN) {\it iniziativa specifica} MOONLIGHT2.


\bibliography{biblio}


\begin{thebibliography}{92}
\ifx \bisbn   \undefined \def \bisbn  #1{ISBN #1}\fi
\ifx \binits  \undefined \def \binits#1{#1}\fi
\ifx \bauthor  \undefined \def \bauthor#1{#1}\fi
\ifx \batitle  \undefined \def \batitle#1{#1}\fi
\ifx \bjtitle  \undefined \def \bjtitle#1{#1}\fi
\ifx \bvolume  \undefined \def \bvolume#1{\textbf{#1}}\fi
\ifx \byear  \undefined \def \byear#1{#1}\fi
\ifx \bissue  \undefined \def \bissue#1{#1}\fi
\ifx \bfpage  \undefined \def \bfpage#1{#1}\fi
\ifx \blpage  \undefined \def \blpage #1{#1}\fi
\ifx \burl  \undefined \def \burl#1{\textsf{#1}}\fi
\ifx \doiurl  \undefined \def \doiurl#1{\url{https://doi.org/#1}}\fi
\ifx \betal  \undefined \def \betal{\textit{et al.}}\fi
\ifx \binstitute  \undefined \def \binstitute#1{#1}\fi
\ifx \binstitutionaled  \undefined \def \binstitutionaled#1{#1}\fi
\ifx \bctitle  \undefined \def \bctitle#1{#1}\fi
\ifx \beditor  \undefined \def \beditor#1{#1}\fi
\ifx \bpublisher  \undefined \def \bpublisher#1{#1}\fi
\ifx \bbtitle  \undefined \def \bbtitle#1{#1}\fi
\ifx \bedition  \undefined \def \bedition#1{#1}\fi
\ifx \bseriesno  \undefined \def \bseriesno#1{#1}\fi
\ifx \blocation  \undefined \def \blocation#1{#1}\fi
\ifx \bsertitle  \undefined \def \bsertitle#1{#1}\fi
\ifx \bsnm \undefined \def \bsnm#1{#1}\fi
\ifx \bsuffix \undefined \def \bsuffix#1{#1}\fi
\ifx \bparticle \undefined \def \bparticle#1{#1}\fi
\ifx \barticle \undefined \def \barticle#1{#1}\fi
\bibcommenthead
\ifx \bconfdate \undefined \def \bconfdate #1{#1}\fi
\ifx \botherref \undefined \def \botherref #1{#1}\fi
\ifx \url \undefined \def \url#1{\textsf{#1}}\fi
\ifx \bchapter \undefined \def \bchapter#1{#1}\fi
\ifx \bbook \undefined \def \bbook#1{#1}\fi
\ifx \bcomment \undefined \def \bcomment#1{#1}\fi
\ifx \oauthor \undefined \def \oauthor#1{#1}\fi
\ifx \citeauthoryear \undefined \def \citeauthoryear#1{#1}\fi
\ifx \endbibitem  \undefined \def \endbibitem {}\fi
\ifx \bconflocation  \undefined \def \bconflocation#1{#1}\fi
\ifx \arxivurl  \undefined \def \arxivurl#1{\textsf{#1}}\fi
\csname PreBibitemsHook\endcsname

\bibitem[\protect\citeauthoryear{Rubakov}{2014}]{Rubakov:2014jja}
\begin{barticle}
\bauthor{\bsnm{Rubakov}, \binits{V.A.}}:
\batitle{{The Null Energy Condition and its violation}}.
\bjtitle{Phys. Usp.}
\bvolume{57},
\bfpage{128}--\blpage{142}
(\byear{2014})
\doiurl{10.3367/UFNe.0184.201402b.0137}
{\href{https://arxiv.org/abs/1401.4024}{{arXiv:1401.4024}}}
{[hep-th]}
\end{barticle}
\endbibitem

\bibitem[\protect\citeauthoryear{Ijjas and Steinhardt}{2016}]{Ijjas:2016tpn}
\begin{barticle}
\bauthor{\bsnm{Ijjas}, \binits{A.}},
\bauthor{\bsnm{Steinhardt}, \binits{P.J.}}:
\batitle{{Classically stable nonsingular cosmological bounces}}.
\bjtitle{Phys. Rev. Lett.}
\bvolume{117}(\bissue{12}),
\bfpage{121304}
(\byear{2016})
\doiurl{10.1103/PhysRevLett.117.121304}
{\href{https://arxiv.org/abs/1606.08880}{{arXiv:1606.08880}}}
{[gr-qc]}
\end{barticle}
\endbibitem

\bibitem[\protect\citeauthoryear{Ijjas and Steinhardt}{2017}]{Ijjas:2016vtq}
\begin{barticle}
\bauthor{\bsnm{Ijjas}, \binits{A.}},
\bauthor{\bsnm{Steinhardt}, \binits{P.J.}}:
\batitle{{Fully stable cosmological solutions with a non-singular classical bounce}}.
\bjtitle{Phys. Lett. B}
\bvolume{764},
\bfpage{289}--\blpage{294}
(\byear{2017})
\doiurl{10.1016/j.physletb.2016.11.047}
{\href{https://arxiv.org/abs/1609.01253}{{arXiv:1609.01253}}}
{[gr-qc]}
\end{barticle}
\endbibitem

\bibitem[\protect\citeauthoryear{Ijjas et~al.}{2019}]{Ijjas:2018cdm}
\begin{barticle}
\bauthor{\bsnm{Ijjas}, \binits{A.}},
\bauthor{\bsnm{Pretorius}, \binits{F.}},
\bauthor{\bsnm{Steinhardt}, \binits{P.J.}}:
\batitle{{Stability and the Gauge Problem in Non-Perturbative Cosmology}}.
\bjtitle{JCAP}
\bvolume{01},
\bfpage{015}
(\byear{2019})
\doiurl{10.1088/1475-7516/2019/01/015}
{\href{https://arxiv.org/abs/1809.07010}{{arXiv:1809.07010}}}
{[gr-qc]}
\end{barticle}
\endbibitem

\bibitem[\protect\citeauthoryear{Tukhashvili and Steinhardt}{2023}]{Tukhashvili:2023itb}
\begin{barticle}
\bauthor{\bsnm{Tukhashvili}, \binits{G.}},
\bauthor{\bsnm{Steinhardt}, \binits{P.J.}}:
\batitle{{Cosmological Bounces Induced by a Fermion Condensate}}.
\bjtitle{Phys. Rev. Lett.}
\bvolume{131}(\bissue{9}),
\bfpage{091001}
(\byear{2023})
\doiurl{10.1103/PhysRevLett.131.091001}
{\href{https://arxiv.org/abs/2307.16098}{{arXiv:2307.16098}}}
{[gr-qc]}
\end{barticle}
\endbibitem

\bibitem[\protect\citeauthoryear{Bojowald}{2001}]{Bojowald:2001xe}
\begin{barticle}
\bauthor{\bsnm{Bojowald}, \binits{M.}}:
\batitle{{Absence of singularity in loop quantum cosmology}}.
\bjtitle{Phys. Rev. Lett.}
\bvolume{86},
\bfpage{5227}--\blpage{5230}
(\byear{2001})
\doiurl{10.1103/PhysRevLett.86.5227}
{\href{https://arxiv.org/abs/gr-qc/0102069}{{arXiv:gr-qc/0102069}}}
\end{barticle}
\endbibitem

\bibitem[\protect\citeauthoryear{Ashtekar and Singh}{2011}]{Ashtekar:2011ni}
\begin{barticle}
\bauthor{\bsnm{Ashtekar}, \binits{A.}},
\bauthor{\bsnm{Singh}, \binits{P.}}:
\batitle{{Loop Quantum Cosmology: A Status Report}}.
\bjtitle{Class. Quant. Grav.}
\bvolume{28},
\bfpage{213001}
(\byear{2011})
\doiurl{10.1088/0264-9381/28/21/213001}
{\href{https://arxiv.org/abs/1108.0893}{{arXiv:1108.0893}}}
{[gr-qc]}
\end{barticle}
\endbibitem

\bibitem[\protect\citeauthoryear{Bojowald}{2019}]{Bojowald:2018sgf}
\begin{barticle}
\bauthor{\bsnm{Bojowald}, \binits{M.}}:
\batitle{{The BKL scenario, infrared renormalization, and quantum cosmology}}.
\bjtitle{JCAP}
\bvolume{01},
\bfpage{026}
(\byear{2019})
\doiurl{10.1088/1475-7516/2019/01/026}
{\href{https://arxiv.org/abs/1810.00238}{{arXiv:1810.00238}}}
{[gr-qc]}
\end{barticle}
\endbibitem

\bibitem[\protect\citeauthoryear{Carroll}{2019}]{Carroll:2004st}
\begin{bbook}
\bauthor{\bsnm{Carroll}, \binits{S.M.}}:
\bbtitle{Spacetime and Geometry: An Introduction to General Relativity}.
\bpublisher{Cambridge University Press}, \blocation{???}
(\byear{2019}).
\doiurl{10.1017/9781108770385}
\end{bbook}
\endbibitem

\bibitem[\protect\citeauthoryear{Capozziello et~al.}{2015}]{Capozziello:2014bqa}
\begin{barticle}
\bauthor{\bsnm{Capozziello}, \binits{S.}},
\bauthor{\bsnm{Lobo}, \binits{F.S.N.}},
\bauthor{\bsnm{Mimoso}, \binits{J.P.}}:
\batitle{{Generalized energy conditions in Extended Theories of Gravity}}.
\bjtitle{Phys. Rev. D}
\bvolume{91}(\bissue{12}),
\bfpage{124019}
(\byear{2015})
\doiurl{10.1103/PhysRevD.91.124019}
{\href{https://arxiv.org/abs/1407.7293}{{arXiv:1407.7293}}}
{[gr-qc]}
\end{barticle}
\endbibitem

\bibitem[\protect\citeauthoryear{Ijjas et~al.}{2016}]{Ijjas:2016pad}
\begin{barticle}
\bauthor{\bsnm{Ijjas}, \binits{A.}},
\bauthor{\bsnm{Ripley}, \binits{J.}},
\bauthor{\bsnm{Steinhardt}, \binits{P.J.}}:
\batitle{{NEC violation in mimetic cosmology revisited}}.
\bjtitle{Phys. Lett. B}
\bvolume{760},
\bfpage{132}--\blpage{138}
(\byear{2016})
\doiurl{10.1016/j.physletb.2016.06.052}
{\href{https://arxiv.org/abs/1604.08586}{{arXiv:1604.08586}}}
{[gr-qc]}
\end{barticle}
\endbibitem

\bibitem[\protect\citeauthoryear{Taveras}{2008}]{Taveras:2008ke}
\begin{barticle}
\bauthor{\bsnm{Taveras}, \binits{V.}}:
\batitle{{Corrections to the Friedmann Equations from LQG for a Universe with a Free Scalar Field}}.
\bjtitle{Phys. Rev. D}
\bvolume{78},
\bfpage{064072}
(\byear{2008})
\doiurl{10.1103/PhysRevD.78.064072}
{\href{https://arxiv.org/abs/0807.3325}{{arXiv:0807.3325}}}
{[gr-qc]}
\end{barticle}
\endbibitem

\bibitem[\protect\citeauthoryear{Singh et~al.}{2006}]{Singh:2006im}
\begin{barticle}
\bauthor{\bsnm{Singh}, \binits{P.}},
\bauthor{\bsnm{Vandersloot}, \binits{K.}},
\bauthor{\bsnm{Vereshchagin}, \binits{G.V.}}:
\batitle{{Non-singular bouncing universes in loop quantum cosmology}}.
\bjtitle{Phys. Rev. D}
\bvolume{74},
\bfpage{043510}
(\byear{2006})
\doiurl{10.1103/PhysRevD.74.043510}
{\href{https://arxiv.org/abs/gr-qc/0606032}{{arXiv:gr-qc/0606032}}}
\end{barticle}
\endbibitem

\bibitem[\protect\citeauthoryear{Sami et~al.}{2006}]{Sami:2006wj}
\begin{barticle}
\bauthor{\bsnm{Sami}, \binits{M.}},
\bauthor{\bsnm{Singh}, \binits{P.}},
\bauthor{\bsnm{Tsujikawa}, \binits{S.}}:
\batitle{{Avoidance of future singularities in loop quantum cosmology}}.
\bjtitle{Phys. Rev. D}
\bvolume{74},
\bfpage{043514}
(\byear{2006})
\doiurl{10.1103/PhysRevD.74.043514}
{\href{https://arxiv.org/abs/gr-qc/0605113}{{arXiv:gr-qc/0605113}}}
\end{barticle}
\endbibitem

\bibitem[\protect\citeauthoryear{Bojowald}{2012}]{Bojowald:2012xy}
\begin{barticle}
\bauthor{\bsnm{Bojowald}, \binits{M.}}:
\batitle{{Quantum Cosmology: Effective Theory}}.
\bjtitle{Class. Quant. Grav.}
\bvolume{29},
\bfpage{213001}
(\byear{2012})
\doiurl{10.1088/0264-9381/29/21/213001}
{\href{https://arxiv.org/abs/1209.3403}{{arXiv:1209.3403}}}
{[gr-qc]}
\end{barticle}
\endbibitem

\bibitem[\protect\citeauthoryear{Mielczarek and Szydlowski}{2008}]{Mielczarek:2008zz}
\begin{barticle}
\bauthor{\bsnm{Mielczarek}, \binits{J.}},
\bauthor{\bsnm{Szydlowski}, \binits{M.}}:
\batitle{{Emerging singularities in the bouncing loop cosmology}}.
\bjtitle{Phys. Rev. D}
\bvolume{77},
\bfpage{124008}
(\byear{2008})
\doiurl{10.1103/PhysRevD.77.124008}
{\href{https://arxiv.org/abs/0801.1073}{{arXiv:0801.1073}}}
{[gr-qc]}
\end{barticle}
\endbibitem

\bibitem[\protect\citeauthoryear{Li et~al.}{2018a}]{Li:2018opr}
\begin{barticle}
\bauthor{\bsnm{Li}, \binits{B.-F.}},
\bauthor{\bsnm{Singh}, \binits{P.}},
\bauthor{\bsnm{Wang}, \binits{A.}}:
\batitle{{Towards Cosmological Dynamics from Loop Quantum Gravity}}.
\bjtitle{Phys. Rev. D}
\bvolume{97}(\bissue{8}),
\bfpage{084029}
(\byear{2018})
\doiurl{10.1103/PhysRevD.97.084029}
{\href{https://arxiv.org/abs/1801.07313}{{arXiv:1801.07313}}}
{[gr-qc]}
\end{barticle}
\endbibitem

\bibitem[\protect\citeauthoryear{Li et~al.}{2018b}]{Li:2018fco}
\begin{barticle}
\bauthor{\bsnm{Li}, \binits{B.-F.}},
\bauthor{\bsnm{Singh}, \binits{P.}},
\bauthor{\bsnm{Wang}, \binits{A.}}:
\batitle{{Qualitative dynamics and inflationary attractors in loop cosmology}}.
\bjtitle{Phys. Rev. D}
\bvolume{98}(\bissue{6}),
\bfpage{066016}
(\byear{2018})
\doiurl{10.1103/PhysRevD.98.066016}
{\href{https://arxiv.org/abs/1807.05236}{{arXiv:1807.05236}}}
{[gr-qc]}
\end{barticle}
\endbibitem

\bibitem[\protect\citeauthoryear{Li et~al.}{2019}]{Li:2019ipm}
\begin{barticle}
\bauthor{\bsnm{Li}, \binits{B.-F.}},
\bauthor{\bsnm{Singh}, \binits{P.}},
\bauthor{\bsnm{Wang}, \binits{A.}}:
\batitle{{Genericness of pre-inflationary dynamics and probability of the desired slow-roll inflation in modified loop quantum cosmologies}}.
\bjtitle{Phys. Rev. D}
\bvolume{100}(\bissue{6}),
\bfpage{063513}
(\byear{2019})
\doiurl{10.1103/PhysRevD.100.063513}
{\href{https://arxiv.org/abs/1906.01001}{{arXiv:1906.01001}}}
{[gr-qc]}
\end{barticle}
\endbibitem

\bibitem[\protect\citeauthoryear{Hrycyna et~al.}{2009}]{Hrycyna:2008yu}
\begin{barticle}
\bauthor{\bsnm{Hrycyna}, \binits{O.}},
\bauthor{\bsnm{Mielczarek}, \binits{J.}},
\bauthor{\bsnm{Szydlowski}, \binits{M.}}:
\batitle{{Effects of the quantisation ambiguities on the Big Bounce dynamics}}.
\bjtitle{Gen. Rel. Grav.}
\bvolume{41},
\bfpage{1025}--\blpage{1049}
(\byear{2009})
\doiurl{10.1007/s10714-008-0689-2}
{\href{https://arxiv.org/abs/0804.2778}{{arXiv:0804.2778}}}
{[gr-qc]}
\end{barticle}
\endbibitem

\bibitem[\protect\citeauthoryear{Dzierzak et~al.}{2009}]{Dzierzak:2009ip}
\begin{barticle}
\bauthor{\bsnm{Dzierzak}, \binits{P.}},
\bauthor{\bsnm{Malkiewicz}, \binits{P.}},
\bauthor{\bsnm{Piechocki}, \binits{W.}}:
\batitle{{Turning big bang into big bounce. 1. Classical dynamics}}.
\bjtitle{Phys. Rev. D}
\bvolume{80},
\bfpage{104001}
(\byear{2009})
\doiurl{10.1103/PhysRevD.80.104001}
{\href{https://arxiv.org/abs/0907.3436}{{arXiv:0907.3436}}}
{[gr-qc]}
\end{barticle}
\endbibitem

\bibitem[\protect\citeauthoryear{Renevey et~al.}{2022}]{Renevey:2021tmh}
\begin{barticle}
\bauthor{\bsnm{Renevey}, \binits{C.}},
\bauthor{\bsnm{Martineau}, \binits{K.}},
\bauthor{\bsnm{Barrau}, \binits{A.}}:
\batitle{{Cosmological implications of generalized holonomy corrections}}.
\bjtitle{Phys. Rev. D}
\bvolume{105}(\bissue{6}),
\bfpage{063521}
(\byear{2022})
\doiurl{10.1103/PhysRevD.105.063521}
{\href{https://arxiv.org/abs/2109.14400}{{arXiv:2109.14400}}}
{[gr-qc]}
\end{barticle}
\endbibitem

\bibitem[\protect\citeauthoryear{Marto et~al.}{2015}]{Marto:2013soa}
\begin{barticle}
\bauthor{\bsnm{Marto}, \binits{J.}},
\bauthor{\bsnm{Tavakoli}, \binits{Y.}},
\bauthor{\bsnm{Vargas~Moniz}, \binits{P.}}:
\batitle{{Improved dynamics and gravitational collapse of tachyon field coupled with a barotropic fluid}}.
\bjtitle{Int. J. Mod. Phys. D}
\bvolume{24}(\bissue{03}),
\bfpage{1550025}
(\byear{2015})
\doiurl{10.1142/S021827181550025X}
{\href{https://arxiv.org/abs/1308.4953}{{arXiv:1308.4953}}}
{[gr-qc]}
\end{barticle}
\endbibitem

\bibitem[\protect\citeauthoryear{Wilson-Ewing}{2015}]{Wilson-Ewing:2015lia}
\begin{barticle}
\bauthor{\bsnm{Wilson-Ewing}, \binits{E.}}:
\batitle{{Loop quantum cosmology with self-dual variables}}.
\bjtitle{Phys. Rev. D}
\bvolume{92}(\bissue{12}),
\bfpage{123536}
(\byear{2015})
\doiurl{10.1103/PhysRevD.92.123536}
{\href{https://arxiv.org/abs/1503.07855}{{arXiv:1503.07855}}}
{[gr-qc]}
\end{barticle}
\endbibitem

\bibitem[\protect\citeauthoryear{Burger et~al.}{2018}]{Burger:2018hpz}
\begin{barticle}
\bauthor{\bsnm{Burger}, \binits{D.J.}},
\bauthor{\bsnm{Moynihan}, \binits{N.}},
\bauthor{\bsnm{Das}, \binits{S.}},
\bauthor{\bsnm{Shajidul~Haque}, \binits{S.}},
\bauthor{\bsnm{Underwood}, \binits{B.}}:
\batitle{{Towards the Raychaudhuri Equation Beyond General Relativity}}.
\bjtitle{Phys. Rev. D}
\bvolume{98}(\bissue{2}),
\bfpage{024006}
(\byear{2018})
\doiurl{10.1103/PhysRevD.98.024006}
{\href{https://arxiv.org/abs/1802.09499}{{arXiv:1802.09499}}}
{[gr-qc]}
\end{barticle}
\endbibitem

\bibitem[\protect\citeauthoryear{Nojiri and Odintsov}{2006}]{Nojiri:2006ri}
\begin{barticle}
\bauthor{\bsnm{Nojiri}, \binits{S.}},
\bauthor{\bsnm{Odintsov}, \binits{S.D.}}:
\batitle{{Introduction to modified gravity and gravitational alternative for dark energy}}.
\bjtitle{eConf}
\bvolume{C0602061},
\bfpage{06}
(\byear{2006})
\doiurl{10.1142/S0219887807001928}
{\href{https://arxiv.org/abs/hep-th/0601213}{{arXiv:hep-th/0601213}}}
\end{barticle}
\endbibitem

\bibitem[\protect\citeauthoryear{Capozziello and Francaviglia}{2008}]{Capozziello:2007ec}
\begin{barticle}
\bauthor{\bsnm{Capozziello}, \binits{S.}},
\bauthor{\bsnm{Francaviglia}, \binits{M.}}:
\batitle{{Extended Theories of Gravity and their Cosmological and Astrophysical Applications}}.
\bjtitle{Gen. Rel. Grav.}
\bvolume{40},
\bfpage{357}--\blpage{420}
(\byear{2008})
\doiurl{10.1007/s10714-007-0551-y}
{\href{https://arxiv.org/abs/0706.1146}{{arXiv:0706.1146}}}
{[astro-ph]}
\end{barticle}
\endbibitem

\bibitem[\protect\citeauthoryear{Sotiriou and Faraoni}{2010}]{Sotiriou:2008rp}
\begin{barticle}
\bauthor{\bsnm{Sotiriou}, \binits{T.P.}},
\bauthor{\bsnm{Faraoni}, \binits{V.}}:
\batitle{{f(R) Theories Of Gravity}}.
\bjtitle{Rev. Mod. Phys.}
\bvolume{82},
\bfpage{451}--\blpage{497}
(\byear{2010})
\doiurl{10.1103/RevModPhys.82.451}
{\href{https://arxiv.org/abs/0805.1726}{{arXiv:0805.1726}}}
{[gr-qc]}
\end{barticle}
\endbibitem

\bibitem[\protect\citeauthoryear{Capozziello and De~Laurentis}{2011}]{Capozziello:2011et}
\begin{barticle}
\bauthor{\bsnm{Capozziello}, \binits{S.}},
\bauthor{\bsnm{De~Laurentis}, \binits{M.}}:
\batitle{{Extended Theories of Gravity}}.
\bjtitle{Phys. Rept.}
\bvolume{509},
\bfpage{167}--\blpage{321}
(\byear{2011})
\doiurl{10.1016/j.physrep.2011.09.003}
{\href{https://arxiv.org/abs/1108.6266}{{arXiv:1108.6266}}}
{[gr-qc]}
\end{barticle}
\endbibitem

\bibitem[\protect\citeauthoryear{Clifton et~al.}{2012}]{Clifton:2011jh}
\begin{barticle}
\bauthor{\bsnm{Clifton}, \binits{T.}},
\bauthor{\bsnm{Ferreira}, \binits{P.G.}},
\bauthor{\bsnm{Padilla}, \binits{A.}},
\bauthor{\bsnm{Skordis}, \binits{C.}}:
\batitle{{Modified Gravity and Cosmology}}.
\bjtitle{Phys. Rept.}
\bvolume{513},
\bfpage{1}--\blpage{189}
(\byear{2012})
\doiurl{10.1016/j.physrep.2012.01.001}
{\href{https://arxiv.org/abs/1106.2476}{{arXiv:1106.2476}}}
{[astro-ph.CO]}
\end{barticle}
\endbibitem

\bibitem[\protect\citeauthoryear{Nojiri et~al.}{2017}]{Nojiri:2017ncd}
\begin{barticle}
\bauthor{\bsnm{Nojiri}, \binits{S.}},
\bauthor{\bsnm{Odintsov}, \binits{S.D.}},
\bauthor{\bsnm{Oikonomou}, \binits{V.K.}}:
\batitle{{Modified Gravity Theories on a Nutshell: Inflation, Bounce and Late-time Evolution}}.
\bjtitle{Phys. Rept.}
\bvolume{692},
\bfpage{1}--\blpage{104}
(\byear{2017})
\doiurl{10.1016/j.physrep.2017.06.001}
{\href{https://arxiv.org/abs/1705.11098}{{arXiv:1705.11098}}}
{[gr-qc]}
\end{barticle}
\endbibitem

\bibitem[\protect\citeauthoryear{Heisenberg}{2019}]{Heisenberg:2018vsk}
\begin{barticle}
\bauthor{\bsnm{Heisenberg}, \binits{L.}}:
\batitle{{A systematic approach to generalisations of General Relativity and their cosmological implications}}.
\bjtitle{Phys. Rept.}
\bvolume{796},
\bfpage{1}--\blpage{113}
(\byear{2019})
\doiurl{10.1016/j.physrep.2018.11.006}
{\href{https://arxiv.org/abs/1807.01725}{{arXiv:1807.01725}}}
{[gr-qc]}
\end{barticle}
\endbibitem

\bibitem[\protect\citeauthoryear{Brandenberger}{2009}]{Brandenberger:2009yt}
\begin{barticle}
\bauthor{\bsnm{Brandenberger}, \binits{R.}}:
\batitle{{Matter Bounce in Horava-Lifshitz Cosmology}}.
\bjtitle{Phys. Rev. D}
\bvolume{80},
\bfpage{043516}
(\byear{2009})
\doiurl{10.1103/PhysRevD.80.043516}
{\href{https://arxiv.org/abs/0904.2835}{{arXiv:0904.2835}}}
{[hep-th]}
\end{barticle}
\endbibitem

\bibitem[\protect\citeauthoryear{Brandenberger and Peter}{2017}]{Brandenberger:2016vhg}
\begin{barticle}
\bauthor{\bsnm{Brandenberger}, \binits{R.}},
\bauthor{\bsnm{Peter}, \binits{P.}}:
\batitle{{Bouncing Cosmologies: Progress and Problems}}.
\bjtitle{Found. Phys.}
\bvolume{47}(\bissue{6}),
\bfpage{797}--\blpage{850}
(\byear{2017})
\doiurl{10.1007/s10701-016-0057-0}
{\href{https://arxiv.org/abs/1603.05834}{{arXiv:1603.05834}}}
{[hep-th]}
\end{barticle}
\endbibitem

\bibitem[\protect\citeauthoryear{Biswas et~al.}{2010}]{Biswas:2010zk}
\begin{barticle}
\bauthor{\bsnm{Biswas}, \binits{T.}},
\bauthor{\bsnm{Koivisto}, \binits{T.}},
\bauthor{\bsnm{Mazumdar}, \binits{A.}}:
\batitle{{Towards a resolution of the cosmological singularity in non-local higher derivative theories of gravity}}.
\bjtitle{JCAP}
\bvolume{11},
\bfpage{008}
(\byear{2010})
\doiurl{10.1088/1475-7516/2010/11/008}
{\href{https://arxiv.org/abs/1005.0590}{{arXiv:1005.0590}}}
{[hep-th]}
\end{barticle}
\endbibitem

\bibitem[\protect\citeauthoryear{Battefeld and Peter}{2015}]{Battefeld:2014uga}
\begin{barticle}
\bauthor{\bsnm{Battefeld}, \binits{D.}},
\bauthor{\bsnm{Peter}, \binits{P.}}:
\batitle{{A Critical Review of Classical Bouncing Cosmologies}}.
\bjtitle{Phys. Rept.}
\bvolume{571},
\bfpage{1}--\blpage{66}
(\byear{2015})
\doiurl{10.1016/j.physrep.2014.12.004}
{\href{https://arxiv.org/abs/1406.2790}{{arXiv:1406.2790}}}
{[astro-ph.CO]}
\end{barticle}
\endbibitem

\bibitem[\protect\citeauthoryear{Cai et~al.}{2014}]{Cai:2014xxa}
\begin{barticle}
\bauthor{\bsnm{Cai}, \binits{Y.-F.}},
\bauthor{\bsnm{Quintin}, \binits{J.}},
\bauthor{\bsnm{Saridakis}, \binits{E.N.}},
\bauthor{\bsnm{Wilson-Ewing}, \binits{E.}}:
\batitle{{Nonsingular bouncing cosmologies in light of BICEP2}}.
\bjtitle{JCAP}
\bvolume{07},
\bfpage{033}
(\byear{2014})
\doiurl{10.1088/1475-7516/2014/07/033}
{\href{https://arxiv.org/abs/1404.4364}{{arXiv:1404.4364}}}
{[astro-ph.CO]}
\end{barticle}
\endbibitem

\bibitem[\protect\citeauthoryear{Cognola et~al.}{2016}]{Cognola:2016gjy}
\begin{barticle}
\bauthor{\bsnm{Cognola}, \binits{G.}},
\bauthor{\bsnm{Myrzakulov}, \binits{R.}},
\bauthor{\bsnm{Sebastiani}, \binits{L.}},
\bauthor{\bsnm{Vagnozzi}, \binits{S.}},
\bauthor{\bsnm{Zerbini}, \binits{S.}}:
\batitle{{Covariant Ho\v{r}ava-like and mimetic Horndeski gravity: cosmological solutions and perturbations}}.
\bjtitle{Class. Quant. Grav.}
\bvolume{33}(\bissue{22}),
\bfpage{225014}
(\byear{2016})
\doiurl{10.1088/0264-9381/33/22/225014}
{\href{https://arxiv.org/abs/1601.00102}{{arXiv:1601.00102}}}
{[gr-qc]}
\end{barticle}
\endbibitem

\bibitem[\protect\citeauthoryear{Ijjas}{2018}]{Ijjas:2017pei}
\begin{barticle}
\bauthor{\bsnm{Ijjas}, \binits{A.}}:
\batitle{{Space-time slicing in Horndeski theories and its implications for non-singular bouncing solutions}}.
\bjtitle{JCAP}
\bvolume{02},
\bfpage{007}
(\byear{2018})
\doiurl{10.1088/1475-7516/2018/02/007}
{\href{https://arxiv.org/abs/1710.05990}{{arXiv:1710.05990}}}
{[gr-qc]}
\end{barticle}
\endbibitem

\bibitem[\protect\citeauthoryear{Kolevatov et~al.}{2017}]{Kolevatov:2017voe}
\begin{barticle}
\bauthor{\bsnm{Kolevatov}, \binits{R.}},
\bauthor{\bsnm{Mironov}, \binits{S.}},
\bauthor{\bsnm{Sukhov}, \binits{N.}},
\bauthor{\bsnm{Volkova}, \binits{V.}}:
\batitle{{Cosmological bounce and Genesis beyond Horndeski}}.
\bjtitle{JCAP}
\bvolume{08},
\bfpage{038}
(\byear{2017})
\doiurl{10.1088/1475-7516/2017/08/038}
{\href{https://arxiv.org/abs/1705.06626}{{arXiv:1705.06626}}}
{[hep-th]}
\end{barticle}
\endbibitem

\bibitem[\protect\citeauthoryear{de~Cesare}{2019}]{deCesare:2018cts}
\begin{barticle}
\bauthor{\bsnm{Cesare}, \binits{M.}}:
\batitle{{Limiting curvature mimetic gravity for group field theory condensates}}.
\bjtitle{Phys. Rev. D}
\bvolume{99}(\bissue{6}),
\bfpage{063505}
(\byear{2019})
\doiurl{10.1103/PhysRevD.99.063505}
{\href{https://arxiv.org/abs/1812.06171}{{arXiv:1812.06171}}}
{[gr-qc]}
\end{barticle}
\endbibitem

\bibitem[\protect\citeauthoryear{Mironov et~al.}{2018}]{Mironov:2018oec}
\begin{barticle}
\bauthor{\bsnm{Mironov}, \binits{S.}},
\bauthor{\bsnm{Rubakov}, \binits{V.}},
\bauthor{\bsnm{Volkova}, \binits{V.}}:
\batitle{{Bounce beyond Horndeski with GR asymptotics and $\gamma$-crossing}}.
\bjtitle{JCAP}
\bvolume{10},
\bfpage{050}
(\byear{2018})
\doiurl{10.1088/1475-7516/2018/10/050}
{\href{https://arxiv.org/abs/1807.08361}{{arXiv:1807.08361}}}
{[hep-th]}
\end{barticle}
\endbibitem

\bibitem[\protect\citeauthoryear{Mironov et~al.}{2019}]{Mironov:2019qjt}
\begin{barticle}
\bauthor{\bsnm{Mironov}, \binits{S.}},
\bauthor{\bsnm{Rubakov}, \binits{V.}},
\bauthor{\bsnm{Volkova}, \binits{V.}}:
\batitle{{Genesis with general relativity asymptotics in beyond Horndeski theory}}.
\bjtitle{Phys. Rev. D}
\bvolume{100}(\bissue{8}),
\bfpage{083521}
(\byear{2019})
\doiurl{10.1103/PhysRevD.100.083521}
{\href{https://arxiv.org/abs/1905.06249}{{arXiv:1905.06249}}}
{[hep-th]}
\end{barticle}
\endbibitem

\bibitem[\protect\citeauthoryear{Mironov et~al.}{2020}]{Mironov:2019mye}
\begin{barticle}
\bauthor{\bsnm{Mironov}, \binits{S.}},
\bauthor{\bsnm{Rubakov}, \binits{V.}},
\bauthor{\bsnm{Volkova}, \binits{V.}}:
\batitle{{Subluminal cosmological bounce beyond Horndeski}}.
\bjtitle{JCAP}
\bvolume{05},
\bfpage{024}
(\byear{2020})
\doiurl{10.1088/1475-7516/2020/05/024}
{\href{https://arxiv.org/abs/1910.07019}{{arXiv:1910.07019}}}
{[hep-th]}
\end{barticle}
\endbibitem

\bibitem[\protect\citeauthoryear{Polarski et~al.}{2022}]{Polarski:2021azv}
\begin{barticle}
\bauthor{\bsnm{Polarski}, \binits{D.}},
\bauthor{\bsnm{Starobinsky}, \binits{A.A.}},
\bauthor{\bsnm{Verbin}, \binits{Y.}}:
\batitle{{Bouncing cosmological isotropic solutions in scalar-tensor gravity}}.
\bjtitle{JCAP}
\bvolume{01}(\bissue{01}),
\bfpage{052}
(\byear{2022})
\doiurl{10.1088/1475-7516/2022/01/052}
{\href{https://arxiv.org/abs/2111.07319}{{arXiv:2111.07319}}}
{[gr-qc]}
\end{barticle}
\endbibitem

\bibitem[\protect\citeauthoryear{Odintsov and Oikonomou}{2015}]{Odintsov:2015zua}
\begin{barticle}
\bauthor{\bsnm{Odintsov}, \binits{S.D.}},
\bauthor{\bsnm{Oikonomou}, \binits{V.K.}}:
\batitle{{$\Lambda$CDM Bounce Cosmology without $\Lambda$CDM: the case of modified gravity}}.
\bjtitle{Phys. Rev. D}
\bvolume{91}(\bissue{6}),
\bfpage{064036}
(\byear{2015})
\doiurl{10.1103/PhysRevD.91.064036}
{\href{https://arxiv.org/abs/1502.06125}{{arXiv:1502.06125}}}
{[gr-qc]}
\end{barticle}
\endbibitem

\bibitem[\protect\citeauthoryear{Nojiri et~al.}{2019}]{Nojiri:2019lqw}
\begin{barticle}
\bauthor{\bsnm{Nojiri}, \binits{S.}},
\bauthor{\bsnm{Odintsov}, \binits{S.D.}},
\bauthor{\bsnm{Oikonomou}, \binits{V.K.}},
\bauthor{\bsnm{Paul}, \binits{T.}}:
\batitle{{Nonsingular bounce cosmology from Lagrange multiplier $F(R)$ gravity}}.
\bjtitle{Phys. Rev. D}
\bvolume{100}(\bissue{8}),
\bfpage{084056}
(\byear{2019})
\doiurl{10.1103/PhysRevD.100.084056}
{\href{https://arxiv.org/abs/1910.03546}{{arXiv:1910.03546}}}
{[gr-qc]}
\end{barticle}
\endbibitem

\bibitem[\protect\citeauthoryear{Bel and Zia}{1985}]{Bel:1985zz}
\begin{barticle}
\bauthor{\bsnm{Bel}, \binits{L.}},
\bauthor{\bsnm{Zia}, \binits{H.S.}}:
\batitle{{Regular reduction of relativistic theories of gravitation with a quadratic Lagrangian}}.
\bjtitle{Phys. Rev. D}
\bvolume{32},
\bfpage{3128}--\blpage{3135}
(\byear{1985})
\doiurl{10.1103/PhysRevD.32.3128}
\end{barticle}
\endbibitem

\bibitem[\protect\citeauthoryear{Simon}{1990}]{Simon:1990ic}
\begin{barticle}
\bauthor{\bsnm{Simon}, \binits{J.Z.}}:
\batitle{{Higher Derivative Lagrangians, Nonlocality, Problems and Solutions}}.
\bjtitle{Phys. Rev. D}
\bvolume{41},
\bfpage{3720}
(\byear{1990})
\doiurl{10.1103/PhysRevD.41.3720}
\end{barticle}
\endbibitem

\bibitem[\protect\citeauthoryear{Sotiriou}{2009}]{Sotiriou:2008ya}
\begin{barticle}
\bauthor{\bsnm{Sotiriou}, \binits{T.P.}}:
\batitle{{Covariant Effective Action for Loop Quantum Cosmology from Order Reduction}}.
\bjtitle{Phys. Rev. D}
\bvolume{79},
\bfpage{044035}
(\byear{2009})
\doiurl{10.1103/PhysRevD.79.044035}
{\href{https://arxiv.org/abs/0811.1799}{{arXiv:0811.1799}}}
{[gr-qc]}
\end{barticle}
\endbibitem

\bibitem[\protect\citeauthoryear{Terrucha et~al.}{2019}]{Terrucha:2019jpm}
\begin{barticle}
\bauthor{\bsnm{Terrucha}, \binits{I.}},
\bauthor{\bsnm{Vernieri}, \binits{D.}},
\bauthor{\bsnm{Lemos}, \binits{J.P.S.}}:
\batitle{{Covariant action for bouncing cosmologies in modified Gauss\textendash{}Bonnet gravity}}.
\bjtitle{Annals Phys.}
\bvolume{404},
\bfpage{39}--\blpage{46}
(\byear{2019})
\doiurl{10.1016/j.aop.2019.02.010}
{\href{https://arxiv.org/abs/1904.00260}{{arXiv:1904.00260}}}
{[gr-qc]}
\end{barticle}
\endbibitem

\bibitem[\protect\citeauthoryear{Barros et~al.}{2020}]{Barros:2019pvc}
\begin{barticle}
\bauthor{\bsnm{Barros}, \binits{B.J.}},
\bauthor{\bsnm{Teixeira}, \binits{E.M.}},
\bauthor{\bsnm{Vernieri}, \binits{D.}}:
\batitle{{Bouncing cosmology in $f(R,\mathcal{G})$ gravity by order reduction}}.
\bjtitle{Annals Phys.}
\bvolume{419},
\bfpage{168231}
(\byear{2020})
\doiurl{10.1016/j.aop.2020.168231}
{\href{https://arxiv.org/abs/1907.11732}{{arXiv:1907.11732}}}
{[gr-qc]}
\end{barticle}
\endbibitem

\bibitem[\protect\citeauthoryear{Bajardi et~al.}{2020}]{Bajardi:2020fxh}
\begin{barticle}
\bauthor{\bsnm{Bajardi}, \binits{F.}},
\bauthor{\bsnm{Vernieri}, \binits{D.}},
\bauthor{\bsnm{Capozziello}, \binits{S.}}:
\batitle{{Bouncing Cosmology in f(Q) Symmetric Teleparallel Gravity}}.
\bjtitle{Eur. Phys. J. Plus}
\bvolume{135}(\bissue{11}),
\bfpage{912}
(\byear{2020})
\doiurl{10.1140/epjp/s13360-020-00918-3}
{\href{https://arxiv.org/abs/2011.01248}{{arXiv:2011.01248}}}
{[gr-qc]}
\end{barticle}
\endbibitem

\bibitem[\protect\citeauthoryear{Miranda et~al.}{2021}]{Miranda:2021oig}
\begin{barticle}
\bauthor{\bsnm{Miranda}, \binits{M.}},
\bauthor{\bsnm{Vernieri}, \binits{D.}},
\bauthor{\bsnm{Capozziello}, \binits{S.}},
\bauthor{\bsnm{Lobo}, \binits{F.S.N.}}:
\batitle{{Effective actions for loop quantum cosmology in fourth-order gravity}}.
\bjtitle{Eur. Phys. J. C}
\bvolume{81}(\bissue{11}),
\bfpage{975}
(\byear{2021})
\doiurl{10.1140/epjc/s10052-021-09767-5}
{\href{https://arxiv.org/abs/2107.07777}{{arXiv:2107.07777}}}
{[gr-qc]}
\end{barticle}
\endbibitem

\bibitem[\protect\citeauthoryear{Ribeiro et~al.}{2023}]{Ribeiro:2021gds}
\begin{barticle}
\bauthor{\bsnm{Ribeiro}, \binits{A.R.}},
\bauthor{\bsnm{Vernieri}, \binits{D.}},
\bauthor{\bsnm{Lobo}, \binits{F.S.N.}}:
\batitle{{Effective f(R) Actions for Modified Loop Quantum Cosmologies via Order Reduction}}.
\bjtitle{Universe}
\bvolume{9}(\bissue{4}),
\bfpage{181}
(\byear{2023})
\doiurl{10.3390/universe9040181}
{\href{https://arxiv.org/abs/2104.12283}{{arXiv:2104.12283}}}
{[gr-qc]}
\end{barticle}
\endbibitem

\bibitem[\protect\citeauthoryear{Allemandi et~al.}{2006}]{Allemandi:2004yx}
\begin{barticle}
\bauthor{\bsnm{Allemandi}, \binits{G.}},
\bauthor{\bsnm{Capone}, \binits{M.}},
\bauthor{\bsnm{Capozziello}, \binits{S.}},
\bauthor{\bsnm{Francaviglia}, \binits{M.}}:
\batitle{{Conformal aspects of Palatini approach in extended theories of gravity}}.
\bjtitle{Gen. Rel. Grav.}
\bvolume{38},
\bfpage{33}--\blpage{60}
(\byear{2006})
\doiurl{10.1007/s10714-005-0208-7}
{\href{https://arxiv.org/abs/hep-th/0409198}{{arXiv:hep-th/0409198}}}
\end{barticle}
\endbibitem

\bibitem[\protect\citeauthoryear{Olmo and Singh}{2009}]{Olmo:2008nf}
\begin{barticle}
\bauthor{\bsnm{Olmo}, \binits{G.J.}},
\bauthor{\bsnm{Singh}, \binits{P.}}:
\batitle{{Effective Action for Loop Quantum Cosmology a la Palatini}}.
\bjtitle{JCAP}
\bvolume{01},
\bfpage{030}
(\byear{2009})
\doiurl{10.1088/1475-7516/2009/01/030}
{\href{https://arxiv.org/abs/0806.2783}{{arXiv:0806.2783}}}
{[gr-qc]}
\end{barticle}
\endbibitem

\bibitem[\protect\citeauthoryear{Barragan et~al.}{2009}]{Barragan:2009sq}
\begin{barticle}
\bauthor{\bsnm{Barragan}, \binits{C.}},
\bauthor{\bsnm{Olmo}, \binits{G.J.}},
\bauthor{\bsnm{Sanchis-Alepuz}, \binits{H.}}:
\batitle{{Bouncing Cosmologies in Palatini f(R) Gravity}}.
\bjtitle{Phys. Rev. D}
\bvolume{80},
\bfpage{024016}
(\byear{2009})
\doiurl{10.1103/PhysRevD.80.024016}
{\href{https://arxiv.org/abs/0907.0318}{{arXiv:0907.0318}}}
{[gr-qc]}
\end{barticle}
\endbibitem

\bibitem[\protect\citeauthoryear{Delhom et~al.}{2023}]{Delhom:2023xxp}
\begin{barticle}
\bauthor{\bsnm{Delhom}, \binits{A.}},
\bauthor{\bsnm{Olmo}, \binits{G.J.}},
\bauthor{\bsnm{Singh}, \binits{P.}}:
\batitle{{A diffeomorphism invariant family of metric-affine actions for loop cosmologies}}.
\bjtitle{JCAP}
\bvolume{06},
\bfpage{059}
(\byear{2023})
\doiurl{10.1088/1475-7516/2023/06/059}
{\href{https://arxiv.org/abs/2302.04285}{{arXiv:2302.04285}}}
{[gr-qc]}
\end{barticle}
\endbibitem

\bibitem[\protect\citeauthoryear{Garriga and Mukhanov}{1999}]{Garriga:1999vw}
\begin{barticle}
\bauthor{\bsnm{Garriga}, \binits{J.}},
\bauthor{\bsnm{Mukhanov}, \binits{V.F.}}:
\batitle{{Perturbations in k-inflation}}.
\bjtitle{Phys. Lett. B}
\bvolume{458},
\bfpage{219}--\blpage{225}
(\byear{1999})
\doiurl{10.1016/S0370-2693(99)00602-4}
{\href{https://arxiv.org/abs/hep-th/9904176}{{arXiv:hep-th/9904176}}}
\end{barticle}
\endbibitem

\bibitem[\protect\citeauthoryear{Armendariz-Picon et~al.}{2001}]{Armendariz-Picon:2000ulo}
\begin{barticle}
\bauthor{\bsnm{Armendariz-Picon}, \binits{C.}},
\bauthor{\bsnm{Mukhanov}, \binits{V.F.}},
\bauthor{\bsnm{Steinhardt}, \binits{P.J.}}:
\batitle{{Essentials of k essence}}.
\bjtitle{Phys. Rev. D}
\bvolume{63},
\bfpage{103510}
(\byear{2001})
\doiurl{10.1103/PhysRevD.63.103510}
{\href{https://arxiv.org/abs/astro-ph/0006373}{{arXiv:astro-ph/0006373}}}
\end{barticle}
\endbibitem

\bibitem[\protect\citeauthoryear{Copeland et~al.}{2006}]{Copeland:2006wr}
\begin{barticle}
\bauthor{\bsnm{Copeland}, \binits{E.J.}},
\bauthor{\bsnm{Sami}, \binits{M.}},
\bauthor{\bsnm{Tsujikawa}, \binits{S.}}:
\batitle{{Dynamics of dark energy}}.
\bjtitle{Int. J. Mod. Phys. D}
\bvolume{15},
\bfpage{1753}--\blpage{1936}
(\byear{2006})
\doiurl{10.1142/S021827180600942X}
{\href{https://arxiv.org/abs/hep-th/0603057}{{arXiv:hep-th/0603057}}}
\end{barticle}
\endbibitem

\bibitem[\protect\citeauthoryear{Faraoni et~al.}{2023}]{Faraoni:2022gry}
\begin{barticle}
\bauthor{\bsnm{Faraoni}, \binits{V.}},
\bauthor{\bsnm{Giardino}, \binits{S.}},
\bauthor{\bsnm{Giusti}, \binits{A.}},
\bauthor{\bsnm{Vanderwee}, \binits{R.}}:
\batitle{{Scalar field as a perfect fluid: thermodynamics of minimally coupled scalars and Einstein frame scalar-tensor gravity}}.
\bjtitle{Eur. Phys. J. C}
\bvolume{83}(\bissue{1}),
\bfpage{24}
(\byear{2023})
\doiurl{10.1140/epjc/s10052-023-11186-7}
{\href{https://arxiv.org/abs/2208.04051}{{arXiv:2208.04051}}}
{[gr-qc]}
\end{barticle}
\endbibitem

\bibitem[\protect\citeauthoryear{Deffayet et~al.}{2010}]{Deffayet:2010qz}
\begin{barticle}
\bauthor{\bsnm{Deffayet}, \binits{C.}},
\bauthor{\bsnm{Pujolas}, \binits{O.}},
\bauthor{\bsnm{Sawicki}, \binits{I.}},
\bauthor{\bsnm{Vikman}, \binits{A.}}:
\batitle{{Imperfect Dark Energy from Kinetic Gravity Braiding}}.
\bjtitle{JCAP}
\bvolume{10},
\bfpage{026}
(\byear{2010})
\doiurl{10.1088/1475-7516/2010/10/026}
{\href{https://arxiv.org/abs/1008.0048}{{arXiv:1008.0048}}}
{[hep-th]}
\end{barticle}
\endbibitem

\bibitem[\protect\citeauthoryear{Gomes and Amendola}{2014}]{Gomes:2013ema}
\begin{barticle}
\bauthor{\bsnm{Gomes}, \binits{A.R.}},
\bauthor{\bsnm{Amendola}, \binits{L.}}:
\batitle{{Towards scaling cosmological solutions with full coupled Horndeski Lagrangian: the KGB model}}.
\bjtitle{JCAP}
\bvolume{03},
\bfpage{041}
(\byear{2014})
\doiurl{10.1088/1475-7516/2014/03/041}
{\href{https://arxiv.org/abs/1306.3593}{{arXiv:1306.3593}}}
{[astro-ph.CO]}
\end{barticle}
\endbibitem

\bibitem[\protect\citeauthoryear{Horndeski}{1974}]{Horndeski:1974wa}
\begin{barticle}
\bauthor{\bsnm{Horndeski}, \binits{G.W.}}:
\batitle{{Second-order scalar-tensor field equations in a four-dimensional space}}.
\bjtitle{Int. J. Theor. Phys.}
\bvolume{10},
\bfpage{363}--\blpage{384}
(\byear{1974})
\doiurl{10.1007/BF01807638}
\end{barticle}
\endbibitem

\bibitem[\protect\citeauthoryear{Deffayet et~al.}{2011}]{Deffayet:2011gz}
\begin{barticle}
\bauthor{\bsnm{Deffayet}, \binits{C.}},
\bauthor{\bsnm{Gao}, \binits{X.}},
\bauthor{\bsnm{Steer}, \binits{D.A.}},
\bauthor{\bsnm{Zahariade}, \binits{G.}}:
\batitle{{From k-essence to generalised Galileons}}.
\bjtitle{Phys. Rev. D}
\bvolume{84},
\bfpage{064039}
(\byear{2011})
\doiurl{10.1103/PhysRevD.84.064039}
{\href{https://arxiv.org/abs/1103.3260}{{arXiv:1103.3260}}}
{[hep-th]}
\end{barticle}
\endbibitem

\bibitem[\protect\citeauthoryear{Kobayashi et~al.}{2011}]{Kobayashi:2011nu}
\begin{barticle}
\bauthor{\bsnm{Kobayashi}, \binits{T.}},
\bauthor{\bsnm{Yamaguchi}, \binits{M.}},
\bauthor{\bsnm{Yokoyama}, \binits{J.}}:
\batitle{{Generalized G-inflation: Inflation with the most general second-order field equations}}.
\bjtitle{Prog. Theor. Phys.}
\bvolume{126},
\bfpage{511}--\blpage{529}
(\byear{2011})
\doiurl{10.1143/PTP.126.511}
{\href{https://arxiv.org/abs/1105.5723}{{arXiv:1105.5723}}}
{[hep-th]}
\end{barticle}
\endbibitem

\bibitem[\protect\citeauthoryear{Afshordi et~al.}{2007a}]{Afshordi:2006ad}
\begin{barticle}
\bauthor{\bsnm{Afshordi}, \binits{N.}},
\bauthor{\bsnm{Chung}, \binits{D.J.H.}},
\bauthor{\bsnm{Geshnizjani}, \binits{G.}}:
\batitle{{Cuscuton: A Causal Field Theory with an Infinite Speed of Sound}}.
\bjtitle{Phys. Rev. D}
\bvolume{75},
\bfpage{083513}
(\byear{2007})
\doiurl{10.1103/PhysRevD.75.083513}
{\href{https://arxiv.org/abs/hep-th/0609150}{{arXiv:hep-th/0609150}}}
\end{barticle}
\endbibitem

\bibitem[\protect\citeauthoryear{Afshordi et~al.}{2007b}]{Afshordi:2007yx}
\begin{barticle}
\bauthor{\bsnm{Afshordi}, \binits{N.}},
\bauthor{\bsnm{Chung}, \binits{D.J.H.}},
\bauthor{\bsnm{Doran}, \binits{M.}},
\bauthor{\bsnm{Geshnizjani}, \binits{G.}}:
\batitle{{Cuscuton Cosmology: Dark Energy meets Modified Gravity}}.
\bjtitle{Phys. Rev. D}
\bvolume{75},
\bfpage{123509}
(\byear{2007})
\doiurl{10.1103/PhysRevD.75.123509}
{\href{https://arxiv.org/abs/astro-ph/0702002}{{arXiv:astro-ph/0702002}}}
\end{barticle}
\endbibitem

\bibitem[\protect\citeauthoryear{Afshordi}{2009}]{Afshordi:2009tt}
\begin{barticle}
\bauthor{\bsnm{Afshordi}, \binits{N.}}:
\batitle{{Cuscuton and low energy limit of Horava-Lifshitz gravity}}.
\bjtitle{Phys. Rev. D}
\bvolume{80},
\bfpage{081502}
(\byear{2009})
\doiurl{10.1103/PhysRevD.80.081502}
{\href{https://arxiv.org/abs/0907.5201}{{arXiv:0907.5201}}}
{[hep-th]}
\end{barticle}
\endbibitem

\bibitem[\protect\citeauthoryear{Bhattacharyya et~al.}{2018}]{Bhattacharyya:2016mah}
\begin{barticle}
\bauthor{\bsnm{Bhattacharyya}, \binits{J.}},
\bauthor{\bsnm{Coates}, \binits{A.}},
\bauthor{\bsnm{Colombo}, \binits{M.}},
\bauthor{\bsnm{Gumrukcuoglu}, \binits{A.E.}},
\bauthor{\bsnm{Sotiriou}, \binits{T.P.}}:
\batitle{{Revisiting the cuscuton as a Lorentz-violating gravity theory}}.
\bjtitle{Phys. Rev. D}
\bvolume{97}(\bissue{6}),
\bfpage{064020}
(\byear{2018})
\doiurl{10.1103/PhysRevD.97.064020}
{\href{https://arxiv.org/abs/1612.01824}{{arXiv:1612.01824}}}
{[hep-th]}
\end{barticle}
\endbibitem

\bibitem[\protect\citeauthoryear{Iyonaga et~al.}{2018}]{Iyonaga:2018vnu}
\begin{barticle}
\bauthor{\bsnm{Iyonaga}, \binits{A.}},
\bauthor{\bsnm{Takahashi}, \binits{K.}},
\bauthor{\bsnm{Kobayashi}, \binits{T.}}:
\batitle{{Extended Cuscuton: Formulation}}.
\bjtitle{JCAP}
\bvolume{12},
\bfpage{002}
(\byear{2018})
\doiurl{10.1088/1475-7516/2018/12/002}
{\href{https://arxiv.org/abs/1809.10935}{{arXiv:1809.10935}}}
{[gr-qc]}
\end{barticle}
\endbibitem

\bibitem[\protect\citeauthoryear{Quintin and Yoshida}{2020}]{Quintin:2019orx}
\begin{barticle}
\bauthor{\bsnm{Quintin}, \binits{J.}},
\bauthor{\bsnm{Yoshida}, \binits{D.}}:
\batitle{{Cuscuton gravity as a classically stable limiting curvature theory}}.
\bjtitle{JCAP}
\bvolume{02},
\bfpage{016}
(\byear{2020})
\doiurl{10.1088/1475-7516/2020/02/016}
{\href{https://arxiv.org/abs/1911.06040}{{arXiv:1911.06040}}}
{[gr-qc]}
\end{barticle}
\endbibitem

\bibitem[\protect\citeauthoryear{Miranda et~al.}{2022}]{Miranda:2022brj}
\begin{barticle}
\bauthor{\bsnm{Miranda}, \binits{M.}},
\bauthor{\bsnm{Vernieri}, \binits{D.}},
\bauthor{\bsnm{Capozziello}, \binits{S.}},
\bauthor{\bsnm{Faraoni}, \binits{V.}}:
\batitle{{Generalized McVittie geometry in Horndeski gravity with matter}}.
\bjtitle{Phys. Rev. D}
\bvolume{105}(\bissue{12}),
\bfpage{124024}
(\byear{2022})
\doiurl{10.1103/PhysRevD.105.124024}
{\href{https://arxiv.org/abs/2204.09693}{{arXiv:2204.09693}}}
{[gr-qc]}
\end{barticle}
\endbibitem

\bibitem[\protect\citeauthoryear{Miranda et~al.}{2023}]{Miranda:2022wkz}
\begin{barticle}
\bauthor{\bsnm{Miranda}, \binits{M.}},
\bauthor{\bsnm{Vernieri}, \binits{D.}},
\bauthor{\bsnm{Capozziello}, \binits{S.}},
\bauthor{\bsnm{Faraoni}, \binits{V.}}:
\batitle{{Fluid nature constrains Horndeski gravity}}.
\bjtitle{Gen. Rel. Grav.}
\bvolume{55}(\bissue{7}),
\bfpage{84}
(\byear{2023})
\doiurl{10.1007/s10714-023-03128-1}
{\href{https://arxiv.org/abs/2209.02727}{{arXiv:2209.02727}}}
{[gr-qc]}
\end{barticle}
\endbibitem

\bibitem[\protect\citeauthoryear{Creminelli and Vernizzi}{2017}]{Creminelli:2017sry}
\begin{barticle}
\bauthor{\bsnm{Creminelli}, \binits{P.}},
\bauthor{\bsnm{Vernizzi}, \binits{F.}}:
\batitle{{Dark Energy after GW170817 and GRB170817A}}.
\bjtitle{Phys. Rev. Lett.}
\bvolume{119}(\bissue{25}),
\bfpage{251302}
(\byear{2017})
\doiurl{10.1103/PhysRevLett.119.251302}
{\href{https://arxiv.org/abs/1710.05877}{{arXiv:1710.05877}}}
{[astro-ph.CO]}
\end{barticle}
\endbibitem

\bibitem[\protect\citeauthoryear{Baker et~al.}{2017}]{Baker:2017hug}
\begin{barticle}
\bauthor{\bsnm{Baker}, \binits{T.}},
\bauthor{\bsnm{Bellini}, \binits{E.}},
\bauthor{\bsnm{Ferreira}, \binits{P.G.}},
\bauthor{\bsnm{Lagos}, \binits{M.}},
\bauthor{\bsnm{Noller}, \binits{J.}},
\bauthor{\bsnm{Sawicki}, \binits{I.}}:
\batitle{{Strong constraints on cosmological gravity from GW170817 and GRB 170817A}}.
\bjtitle{Phys. Rev. Lett.}
\bvolume{119}(\bissue{25}),
\bfpage{251301}
(\byear{2017})
\doiurl{10.1103/PhysRevLett.119.251301}
{\href{https://arxiv.org/abs/1710.06394}{{arXiv:1710.06394}}}
{[astro-ph.CO]}
\end{barticle}
\endbibitem

\bibitem[\protect\citeauthoryear{Bettoni et~al.}{2017}]{Bettoni:2016mij}
\begin{barticle}
\bauthor{\bsnm{Bettoni}, \binits{D.}},
\bauthor{\bsnm{Ezquiaga}, \binits{J.M.}},
\bauthor{\bsnm{Hinterbichler}, \binits{K.}},
\bauthor{\bsnm{Zumalac\'arregui}, \binits{M.}}:
\batitle{{Speed of Gravitational Waves and the Fate of Scalar-Tensor Gravity}}.
\bjtitle{Phys. Rev. D}
\bvolume{95}(\bissue{8}),
\bfpage{084029}
(\byear{2017})
\doiurl{10.1103/PhysRevD.95.084029}
{\href{https://arxiv.org/abs/1608.01982}{{arXiv:1608.01982}}}
{[gr-qc]}
\end{barticle}
\endbibitem

\bibitem[\protect\citeauthoryear{Andreou et~al.}{2019}]{Andreou:2019ikc}
\begin{barticle}
\bauthor{\bsnm{Andreou}, \binits{N.}},
\bauthor{\bsnm{Franchini}, \binits{N.}},
\bauthor{\bsnm{Ventagli}, \binits{G.}},
\bauthor{\bsnm{Sotiriou}, \binits{T.P.}}:
\batitle{{Spontaneous scalarization in generalised scalar-tensor theory}}.
\bjtitle{Phys. Rev. D}
\bvolume{99}(\bissue{12}),
\bfpage{124022}
(\byear{2019})
\doiurl{10.1103/PhysRevD.99.124022}
{\href{https://arxiv.org/abs/1904.06365}{{arXiv:1904.06365}}}
{[gr-qc]}.
\bcomment{[Erratum: Phys.Rev.D 101, 109903 (2020)]}
\end{barticle}
\endbibitem

\bibitem[\protect\citeauthoryear{Pimentel}{1989}]{Pimentel:1989bm}
\begin{barticle}
\bauthor{\bsnm{Pimentel}, \binits{L.O.}}:
\batitle{{Energy Momentum Tensor in the General Scalar - Tensor Theory}}.
\bjtitle{Class. Quant. Grav.}
\bvolume{6},
\bfpage{263}--\blpage{265}
(\byear{1989})
\doiurl{10.1088/0264-9381/6/12/005}
\end{barticle}
\endbibitem

\bibitem[\protect\citeauthoryear{Pujolas et~al.}{2011}]{Pujolas:2011he}
\begin{barticle}
\bauthor{\bsnm{Pujolas}, \binits{O.}},
\bauthor{\bsnm{Sawicki}, \binits{I.}},
\bauthor{\bsnm{Vikman}, \binits{A.}}:
\batitle{{The Imperfect Fluid behind Kinetic Gravity Braiding}}.
\bjtitle{JHEP}
\bvolume{11},
\bfpage{156}
(\byear{2011})
\doiurl{10.1007/JHEP11(2011)156}
{\href{https://arxiv.org/abs/1103.5360}{{arXiv:1103.5360}}}
{[hep-th]}
\end{barticle}
\endbibitem

\bibitem[\protect\citeauthoryear{Faraoni and Cot\'e}{2018}]{Faraoni:2018qdr}
\begin{barticle}
\bauthor{\bsnm{Faraoni}, \binits{V.}},
\bauthor{\bsnm{Cot\'e}, \binits{J.}}:
\batitle{{Imperfect fluid description of modified gravities}}.
\bjtitle{Phys. Rev. D}
\bvolume{98}(\bissue{8}),
\bfpage{084019}
(\byear{2018})
\doiurl{10.1103/PhysRevD.98.084019}
{\href{https://arxiv.org/abs/1808.02427}{{arXiv:1808.02427}}}
{[gr-qc]}
\end{barticle}
\endbibitem

\bibitem[\protect\citeauthoryear{Giusti et~al.}{2022}]{Giusti:2021sku}
\begin{barticle}
\bauthor{\bsnm{Giusti}, \binits{A.}},
\bauthor{\bsnm{Zentarra}, \binits{S.}},
\bauthor{\bsnm{Heisenberg}, \binits{L.}},
\bauthor{\bsnm{Faraoni}, \binits{V.}}:
\batitle{{First-order thermodynamics of Horndeski gravity}}.
\bjtitle{Phys. Rev. D}
\bvolume{105}(\bissue{12}),
\bfpage{124011}
(\byear{2022})
\doiurl{10.1103/PhysRevD.105.124011}
{\href{https://arxiv.org/abs/2108.10706}{{arXiv:2108.10706}}}
{[gr-qc]}
\end{barticle}
\endbibitem

\bibitem[\protect\citeauthoryear{Miranda et~al.}{2024}]{Miranda:2024dhw}
\begin{botherref}
\oauthor{\bsnm{Miranda}, \binits{M.}},
\oauthor{\bsnm{Giardino}, \binits{S.}},
\oauthor{\bsnm{Giusti}, \binits{A.}},
\oauthor{\bsnm{Heisenberg}, \binits{L.}}:
{First-order thermodynamics of Horndeski cosmology}
(2024)
{\href{https://arxiv.org/abs/2401.10351}{{arXiv:2401.10351}}}
{[gr-qc]}
\end{botherref}
\endbibitem

\bibitem[\protect\citeauthoryear{Gao and Yao}{2020}]{Gao:2019twq}
\begin{barticle}
\bauthor{\bsnm{Gao}, \binits{X.}},
\bauthor{\bsnm{Yao}, \binits{Z.-B.}}:
\batitle{{Spatially covariant gravity theories with two tensorial degrees of freedom: the formalism}}.
\bjtitle{Phys. Rev. D}
\bvolume{101}(\bissue{6}),
\bfpage{064018}
(\byear{2020})
\doiurl{10.1103/PhysRevD.101.064018}
{\href{https://arxiv.org/abs/1910.13995}{{arXiv:1910.13995}}}
{[gr-qc]}
\end{barticle}
\endbibitem

\bibitem[\protect\citeauthoryear{Boruah et~al.}{2018}]{Boruah:2018pvq}
\begin{barticle}
\bauthor{\bsnm{Boruah}, \binits{S.S.}},
\bauthor{\bsnm{Kim}, \binits{H.J.}},
\bauthor{\bsnm{Rouben}, \binits{M.}},
\bauthor{\bsnm{Geshnizjani}, \binits{G.}}:
\batitle{{Cuscuton bounce}}.
\bjtitle{JCAP}
\bvolume{08},
\bfpage{031}
(\byear{2018})
\doiurl{10.1088/1475-7516/2018/08/031}
{\href{https://arxiv.org/abs/1802.06818}{{arXiv:1802.06818}}}
{[gr-qc]}
\end{barticle}
\endbibitem

\bibitem[\protect\citeauthoryear{Ito et~al.}{2019}]{Ito:2019fie}
\begin{barticle}
\bauthor{\bsnm{Ito}, \binits{A.}},
\bauthor{\bsnm{Iyonaga}, \binits{A.}},
\bauthor{\bsnm{Kim}, \binits{S.}},
\bauthor{\bsnm{Soda}, \binits{J.}}:
\batitle{{Dressed power-law inflation with a cuscuton}}.
\bjtitle{Phys. Rev. D}
\bvolume{99}(\bissue{8}),
\bfpage{083502}
(\byear{2019})
\doiurl{10.1103/PhysRevD.99.083502}
{\href{https://arxiv.org/abs/1902.08663}{{arXiv:1902.08663}}}
{[astro-ph.CO]}
\end{barticle}
\endbibitem

\bibitem[\protect\citeauthoryear{Kim and Geshnizjani}{2021}]{Kim:2020iwq}
\begin{barticle}
\bauthor{\bsnm{Kim}, \binits{J.L.}},
\bauthor{\bsnm{Geshnizjani}, \binits{G.}}:
\batitle{{Spectrum of Cuscuton Bounce}}.
\bjtitle{JCAP}
\bvolume{03},
\bfpage{104}
(\byear{2021})
\doiurl{10.1088/1475-7516/2021/03/104}
{\href{https://arxiv.org/abs/2010.06645}{{arXiv:2010.06645}}}
{[gr-qc]}
\end{barticle}
\endbibitem

\bibitem[\protect\citeauthoryear{Bartolo et~al.}{2022}]{Bartolo:2021wpt}
\begin{barticle}
\bauthor{\bsnm{Bartolo}, \binits{N.}},
\bauthor{\bsnm{Ganz}, \binits{A.}},
\bauthor{\bsnm{Matarrese}, \binits{S.}}:
\batitle{{Cuscuton inflation}}.
\bjtitle{JCAP}
\bvolume{05}(\bissue{05}),
\bfpage{008}
(\byear{2022})
\doiurl{10.1088/1475-7516/2022/05/008}
{\href{https://arxiv.org/abs/2111.06794}{{arXiv:2111.06794}}}
{[gr-qc]}
\end{barticle}
\endbibitem

\bibitem[\protect\citeauthoryear{Papallo and Reall}{2017}]{Papallo:2017qvl}
\begin{barticle}
\bauthor{\bsnm{Papallo}, \binits{G.}},
\bauthor{\bsnm{Reall}, \binits{H.S.}}:
\batitle{{On the local well-posedness of Lovelock and Horndeski theories}}.
\bjtitle{Phys. Rev. D}
\bvolume{96}(\bissue{4}),
\bfpage{044019}
(\byear{2017})
\doiurl{10.1103/PhysRevD.96.044019}
{\href{https://arxiv.org/abs/1705.04370}{{arXiv:1705.04370}}}
{[gr-qc]}
\end{barticle}
\endbibitem

\bibitem[\protect\citeauthoryear{Kov\'acs and Reall}{2020}]{Kovacs:2020ywu}
\begin{barticle}
\bauthor{\bsnm{Kov\'acs}, \binits{A.D.}},
\bauthor{\bsnm{Reall}, \binits{H.S.}}:
\batitle{{Well-posed formulation of Lovelock and Horndeski theories}}.
\bjtitle{Phys. Rev. D}
\bvolume{101}(\bissue{12}),
\bfpage{124003}
(\byear{2020})
\doiurl{10.1103/PhysRevD.101.124003}
{\href{https://arxiv.org/abs/2003.08398}{{arXiv:2003.08398}}}
{[gr-qc]}
\end{barticle}
\endbibitem

\end{thebibliography}


\end{document}